\documentclass[conference]{IEEEtran}
\usepackage{tikz}
\usepackage{amsmath}

\usepackage{comment}
\usepackage{amssymb}
\usepackage{booktabs}
\usepackage{pifont}
\usepackage{multirow}
\usepackage{multicol}
\usepackage{paralist}
\usepackage{paralist}
\usepackage{enumitem}
\usepackage{xcolor} %

\usepackage[linesnumbered,ruled]{algorithm2e}

\usepackage{wasysym}

\DeclareMathOperator*{\argmax}{arg\,max}

\mathchardef\mhyphen="2D %

\newcommand{\RR}{\mathbb{R}}
\newcommand{\loss}{\mathcal{L}} %

\newcommand{\Classifier}{C} %

\newcommand{\mainacc}{\mathsf{CA}}
\newcommand{\Detector}{D}%

\newcommand{\asr}{\textsf{ASR}\;}
\newcommand{\asrs}{\textsf{ASRs}\;}

\newcommand{\manypixel}{\textsf{MP}\;}
\newcommand{\manypixels}{\textsf{MPs}}
\usepackage{gensymb}

\usepackage[normalem]{ulem}

\usepackage{array}
\newcolumntype{H}{>{\setbox0=\hbox\bgroup}c<{\egroup}@{}}
\usepackage{nicefrac}

\usepackage{cite}
\usepackage{hyperref}
\usepackage{cleveref}

\ifCLASSINFOpdf
\else
\fi

\ifCLASSOPTIONcompsoc
 \usepackage[caption=false,font=normalsize,labelfont=sf,textfont=sf]{subfig}
\else
 \usepackage[caption=false,font=footnotesize]{subfig}
\fi

\usepackage{url}

\begin{document}
\title{Targeted Physical Evasion Attacks in the Near-Infrared Domain}

\author{\IEEEauthorblockN{Pascal Zimmer}
	\IEEEauthorblockA{Ruhr University Bochum\\
		pascal.zimmer@rub.de}
	\and
	\IEEEauthorblockN{Simon Lachnit}
	\IEEEauthorblockA{Ruhr University Bochum\\
		simon.lachnit@rub.de}
	\and
	\IEEEauthorblockN{Alexander Jan Zielinski}
        \IEEEauthorblockA{Ruhr University Bochum\\
		alexander.zielinski@rub.de}
        \and
	\IEEEauthorblockN{Ghassan Karame}
        \IEEEauthorblockA{Ruhr University Bochum\\
		ghassan@karame.org}}

\pagestyle{plain}

\maketitle
\begin{abstract}
A number of attacks rely on infrared light sources or heat-absorbing material to imperceptibly fool systems into misinterpreting visual input in various image recognition applications. However, almost all existing approaches can only mount untargeted attacks and require heavy optimizations due to the use-case-specific constraints, such as location and shape.

In this paper, we propose a novel, stealthy, and cost-effective attack to generate both \emph{targeted} and \emph{untargeted} adversarial infrared perturbations. By projecting perturbations from a transparent film onto the target object with an off-the-shelf infrared flashlight, our approach is the first to reliably mount laser-free \emph{targeted} attacks in the infrared domain.
Extensive experiments on traffic signs in the digital and physical domains show that our approach is robust and yields higher attack success rates in various attack scenarios across bright lighting conditions, distances, and angles compared to prior work. Equally important, our attack is highly cost-effective, requiring less than US\$50 and a few tens of seconds for deployment. Finally, we propose a novel segmentation-based detection that thwarts our attack with an F1-score of up to $\mathbf{99\%}$.

\end{abstract}

\IEEEpeerreviewmaketitle

\section{Introduction}
\label{sec:intro}

Deep neural networks are known to be susceptible to malicious inputs, which is especially relevant in safety-critical use cases, such as traffic light/sign recognition and facial recognition systems for access control and surveillance. Different attack strategies exist, some of which assume direct model access and enable direct gradient computations, i.e., white-box model~\cite{goodfellowExplainingHarnessingAdversarial2015}, while others are limited to oracle access to a model (black-box model). In the digital domain, the generation of these perturbations is often constrained with an $L_p$ norm, which captures the difference between a benign and malicious image on a pixel level and is an imperceptibility measure for a (human) observer.

Recent real-world attacks exploit the specificities of camera hardware or hide perturbations in inconspicuous phenomena. Popular methods to conceal perturbations consist of the reliance on so-called adversarial patches; these have been shown to be particularly harmful in traffic sign detection~\cite{zhouDeepBillboardSystematicPhysicalworld2020, duanAdversarialCamouflageHiding2020, eykholtRobustPhysicalWorldAttacks2018}, facial recognition systems~\cite{sharifAccessorizeCrimeReal2016, nguyenAdversarialLightProjection2020}, and person detection \cite{tanLegitimateAdversarialPatches2021, thysFoolingAutomatedSurveillance2019}. Adversarial patches are typically static, leave a visible trace behind, and are constrained in their degrees of freedom. Other methods, such as projector-based attacks, exploit the full RGB color space due to their ability to project arbitrary images with close to pixel-wise precision onto a target. 
Projector-based attacks, however, suffer from a major shortcoming, as the illumination required to be projected onto an object is far from being stealthy.  

Other, more recent, attacks exploit the fact that the sensitivity of CMOS sensors often stretches further into the infrared part of the optical light spectrum compared to human vision, opening the door for exploitation in image recognition applications~\cite{zhouInvisibleMaskPractical2018,wangCanSeeLight2021,satoInvisibleReflectionsLeveraging2024}. 
This problem is further exacerbated by the fact that spectral filters are often not installed in modern vehicles due to their high cost and performance overhead \cite{wangCanSeeLight2021}. In these settings, infrared projectors emerge as a robust and precise means to introduce inconspicuous 
pixel-wise modifications. Such projectors are costly, requiring investments of tens of thousands of USD. As an alternative, several recent contributions have overcome the high cost associated with infrared projectors using infrared lasers~\cite{satoInvisibleReflectionsLeveraging2024}, albeit at the expense of precision. More specifically, even though these approaches can cut down costs to just thousands of USD, \emph{they cannot mount attacks targeting specific classes} since their optimization space is limited and can only reduce the model's confidence in correctly classifying input. As such, \emph{they often result in disruptions of service (e.g., flipping the prediction to any different class) but cannot be used to mount sophisticated attacks}, i.e., precise label-flipping. In comparison, the misclassification of a stop sign as a speed limit 50 sign or vice-versa by an autonomous vehicle poses a greater safety hazard than a simple service disruption. %

In this paper, we propose the first practical and robust infrared perturbation approach to mount inconspicuous \emph{targeted} and \emph{untargeted} attacks in the physical world. Our laser-free approach bridges the gap between powerful projector-based attacks and existing solutions by significantly reducing the complexity of the underlying optimization problem. 
To ensure real-world robustness, we opted to account for the spectral shift into the infrared domain (since we cannot exploit the full RGB color space). We incorporated the use of \emph{expectation over transformation}, i.e., EOT~\cite{athalyeSynthesizingRobustAdversarial}, to adapt to various real-world limitations, e.g., stemming from brightness changes, perturbation misalignment, and spatial transformations.

Unlike previous work~\cite{zhuFoolingThermalInfrared2021, zhuInfraredInvisibleClothing2022, weiHOTCOLDBlockFooling2023, weiPhysicallyAdversarialInfrared2023,satoInvisibleReflectionsLeveraging2024}, our approach considerably reduces the real-world constraints on shape and location by mimicking an infrared projector. This allows us to exploit additional degrees of freedom as a means to generate more robust, targeted, and successful perturbations compared to existing approaches. Moreover, contrary to \cite{zhongShadowsCanBe2022}, our model is not restricted to a single (artificial) light source.  
This particularly allows us to capture realistic deployment environments with varying lighting conditions 
and to realize high-accuracy targeted attacks (in addition to the standard untargeted attacks) with a negligible overhead. Namely, our attack is highly cost- and time-effective---incurring an equipment cost of less than US\$50 and only tens of seconds to deploy. 
In summary, our contributions are as follows: 

\begin{description}[leftmargin=0.25cm]
    \item[Novel attack:] We propose a novel approach to generate adversarial infrared perturbations that alleviates many practical constraints in current proposals and can accurately mount both targeted and untargeted attacks (cf.~\Cref{sec:design}).
    \item[Thorough evaluation:] We evaluate and verify our adversarial infrared perturbations in both targeted and untargeted settings in the use cases of traffic sign recognition, i.e., object detection and image classification, in both digital and physical domains. Real-world experiments show that our approach results in attack success rates of up to $100\%$ in various lighting conditions across varying distances and angles, and in a moving vehicle (up to $30$ km/h), underlining the impact on real-world safety in both two-stage (cf. Section~\ref{sec:experiments}) and single-stage architectures (cf. Section~\ref{single-stage-eval}).
    For instance, our proposal improves the attack success rate by up to {$20.47\%$} compared to~\cite{weiPhysicallyAdversarialInfrared2023,weiHOTCOLDBlockFooling2023}, even though~\cite{weiPhysicallyAdversarialInfrared2023} is a white-box method with direct access to model gradients. We achieve this while requiring a considerably lower number of queries, by up to $65\%$ (cf. \Cref{sec:experiments}).
    \item[Countermeasures:] We show that our proposal exhibits significant robustness against state-of-the-art defensive schemes (cf. \Cref{sec:defenses}). To remedy this, we propose a novel segmentation-based detection scheme that is specifically designed to address infrared perturbation attacks on traffic signs. Our experiments show that our defense can thwart infrared perturbation attacks with an F1-score of up to $99\%$.
    \item[Open science:] Our source code and the first open-source infrared traffic sign dataset, which we dub \emph{GTSRB-IR-100}, to aid researchers in conducting real-world evaluations in the near-infrared spectrum (cf. Appendix~\ref{sec:dataset_ir}), is publicly available\footnote{
    \url{https://github.com/RUB-InfSec/infrared_perturbations}
    }. We also responsibly disclosed our findings to Mercedes, Mobileye, Tesla, Sony, and OnSemi.
\end{description}

\section{Background and Related Work}
\label{sec:back}
\noindent\textbf{Vision-based System Architectures: }
Image recognition architectures generally fall into two categories: single-stage and two-stage~\cite{10.1007/978-3-030-58592-1_5}. Single-stage models perform object detection and classification jointly, offering efficiency for tasks with a limited number of classes, but suffer in performance as the number of classes increases. In contrast, two-stage pipelines first detect objects using a single-class detector and then classify each detected region, making them more suitable for handling a large number of classes.

\vspace{0.5 em}\noindent\textbf{Real-World Adversarial Attacks: }
Adversarial patch attacks have been adapted to real-world settings by embedding visible perturbations that mimic plausible scenarios—e.g., shadows, snow, stickers, or weathered signs~\cite{zhongShadowsCanBe2022, duanAdversarialCamouflageHiding2020, zhouDeepBillboardSystematicPhysicalworld2020}. However, these attacks are often easy to spot due to the unnatural appearance of the patterns.

To increase stealth, newer attacks exploit human perceptual limitations and the characteristics of camera sensors. These include perturbations invisible to humans but detectable by cameras, or those injected via the camera pipeline, such as with modulated lighting~\cite{saylesInvisiblePerturbationsPhysical2021}, laser interference~\cite{yanRollingColorsAdversarial}, ultrasonic signals~\cite{jiPoltergeistAcousticAdversarial2021}, or EM interference~\cite{kohlerSignalInjectionAttacks2022}. Others exploit visual illusions, projecting images too briefly for human detection~\cite{nassiPhantomADASSecuring2020}.

A particularly stealthy class of attacks leverages the camera sensor’s sensitivity to infrared (IR) light. These include hidden IR patterns for evading facial recognition~\cite{zhouInvisibleMaskPractical2018} and spoofing traffic signals with IR LEDs~\cite{wangCanSeeLight2021}. More recent work targets traffic sign detection using large, invisible IR laser spots~\cite{satoInvisibleReflectionsLeveraging2024}, combining techniques from visible-light and laser-based attacks~\cite{liLightCanBe2023, huAdversarialLaserSpot2023}. Unfortunately, due to the limited solution space of possible perturbations, \emph{all existing works can only be effective in the untargeted attack setting}.  

An attacker might also resort to jamming the camera to attack the vision system with (in-)visible light. Jamming attacks are less fine-grained and disable the entire vision system. These attacks also require precise, real-time targeting of the camera of a moving vehicle, making it challenging to execute. 

\section{Design Goals \& Approach}

\subsection{System \& Threat Model}
\begin{figure}[t]
     \centering
     \includegraphics[width=0.75\linewidth]{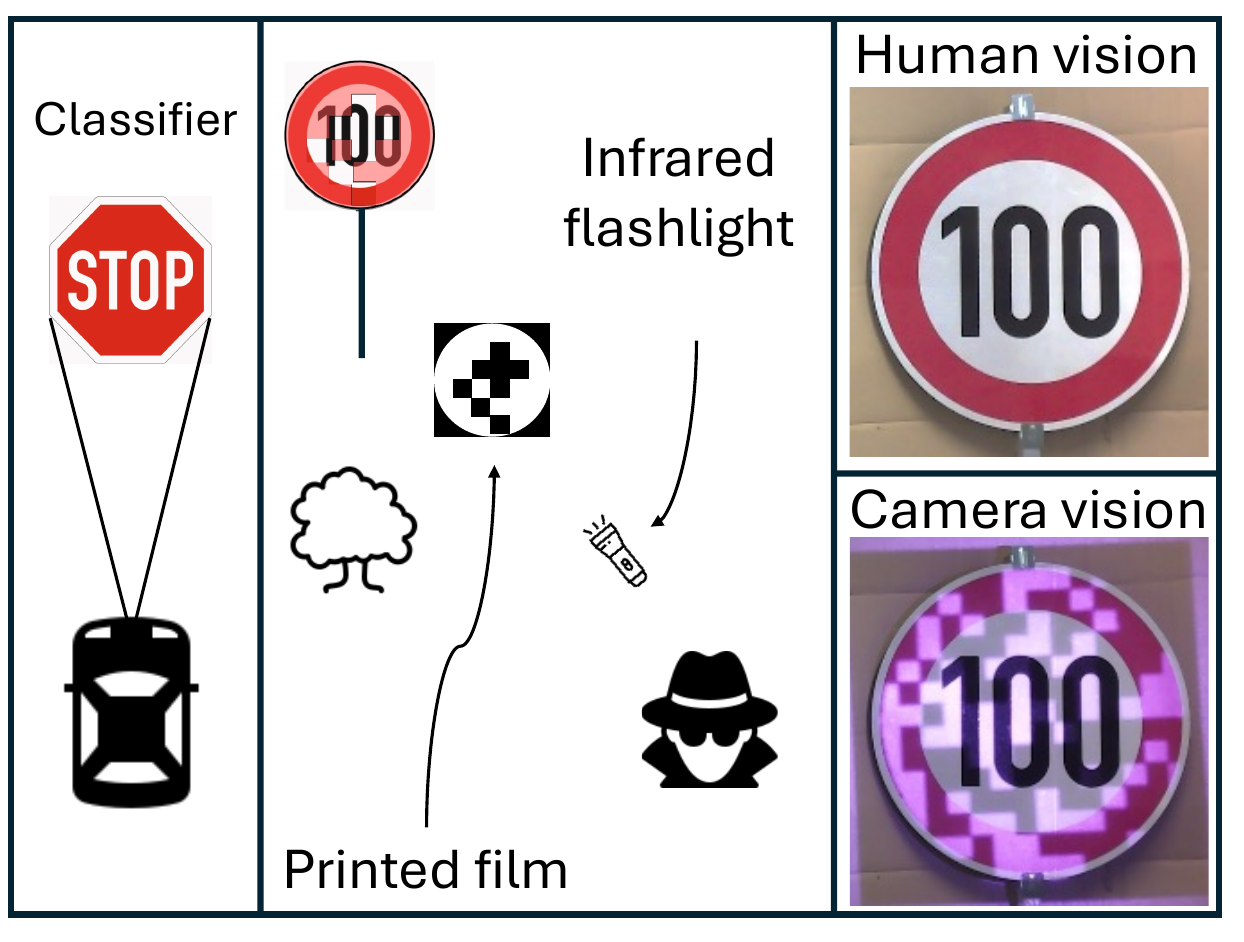}      
  \caption{Overview of our system (icons from~\cite{icon_lic}).}
\label{fig:system_model}
\end{figure}
\label{system_model}
We consider an adversary that is interested in causing vision-based recognition systems used in environments such as autonomous vehicles to output an incorrect prediction by placing an adversarial perturbation on a target object. The adversary is interested in keeping any introduced perturbation invisible to a human observer but clearly observable by a CMOS camera, thereby impacting the image processing pipeline. 

Unless otherwise specified, we focus on the main use case of traffic sign recognition in this work.  
However, we emphasize that our approach is equally applicable to other use cases, such as facial recognition systems. More specifically, we target
both single- and the more challenging two-stage pipelines to ensure that our approach broadly applies to many existing production systems.
The reason that we consider this use case is that it presents an unexplored opportunity for the adversary; in most modern vehicles (e.g., Tesla Model 3), spectral filters are often not installed due to their additional cost, performance overhead, and functional limitations~\cite{wangCanSeeLight2021, 10092278}---preventing the masking of infrared perturbations in modern cars. For instance, spectral filters reduce the perception of a vision system when dealing with low-light conditions, hamper the detection of lane markings~\cite{10092278}, etc.  
Therefore, many vendors explicitly opt for cameras that remain sensitive to NIR, e.g., sensing systems by Mobileye or Mercedes, or are actively developing patents that explicitly consider infrared-sensitive camera systems~\cite{bingleImagingSystemVehicle2011, maoImageMethodImage2018}. 
Moreover, (i) recent camera modules, e.g., e-con Systems STURDeCAM88, are also \emph{not} equipped with infrared filters, and (ii) cutting-edge sensor designs for day-night imaging promise improved color fidelity (mimicking the effect of infrared filters), while maintaining sensitivity to infrared light at night~\cite{maDayNightImagingInfrared2024}. 
 
Throughout this paper, we assume the camera of the attack target lacks infrared filtering and the system relies solely---like recent Tesla models---on vision-based sensing. While some systems may use additional map data, this data is often missing/outdated for new roads, indoor sites, or construction zones. Even when maps are accurate, if a vehicle detects contradicting inputs, e.g., a $20$ km/h sign but the map shows $80$ km/h, it will likely brake. 

Arguably, in this setting, the main goal of the adversary is to manipulate the output of the vision-based recognition system, e.g., by hiding signs or by classifying a stop sign as a speed sign or vice versa (e.g., to target specific manufacturers' processing pipelines~\cite{phanAdversarialImagingPipelines2021}, or to cause harm). 
To achieve these goals, an adversary might resort to three types of attacks: (1) a hiding attack, for which the object detector is fooled to ignore a given object; (2) a weak untargeted attack, for which an image classifier is fooled to output \emph{any} class different from the ground truth, or (3) sophisticated targeted attacks that purposefully intend to fool the classifier into outputting a \emph{specific} target class that is different from the ground truth (e.g.,  misclassify all speed signs as a stop sign). 

Unlike previous work in this area~\cite{satoInvisibleReflectionsLeveraging2024, weiHOTCOLDBlockFooling2023, 287105, weiPhysicallyAdversarialInfrared2023}, we primarily focus on the challenging targeted attack setting; we, nevertheless, also analyze and evaluate the effectiveness of our approach in the untargeted setting.
In contrast to camera jamming, our infrared perturbation selectively misclassifies specific signs without disrupting functionality. 

To achieve these goals, we assume that the adversary has black-box score-based oracle access to the target image classifier---in which an adversary can only observe the output probabilities after supplying a controlled input. That is, we assume that the adversary neither knows the model weights, architecture, nor has access to the exact training data. This mimics a realistic setting where the adversary can only interact, e.g., with the classifier of a vehicle through a debug interface, but does not have access to the full classifier~\cite{huAdversarialLaserSpot2023, jiPoltergeistAcousticAdversarial2021,  liLightCanBe2023,satoInvisibleReflectionsLeveraging2024,weiHOTCOLDBlockFooling2023,zhongShadowsCanBe2022}.\footnote{The adversary can be a user themselves to interact with the classifier to directly observe how a manipulated sign is perceived on a car's dashboard.} 

\subsection{Design Criteria}

To make the physical adversarial perturbations practical, our design seeks to achieve the following criteria.

\vspace{0.5 em}\noindent \textbf{Targeted, Untargeted \& Hide Attacks.} Adversarial perturbations should effectively realize targeted, untargeted, and hide attacks. In most practical deployments, hide and untargeted attacks result in service disruptions (e.g., by hiding an object or preventing correct classification). Targeted attacks, on the other hand, are more powerful as they enable the adversary to cause specific damage, such as causing autonomous vehicles to increase their speed at stop signs. 

\vspace{0.5 em}\noindent \textbf{System and Camera-Agnostic.} Since it is not feasible for an attacker to obtain the images captured from the camera of an approaching vehicle in real-time, the crafted adversarial perturbations should not be specific to a given (camera) system and should be transferable across various target classifiers. 

\vspace{0.5 em}\noindent \textbf{Scene-Agnostic.} The adversary may have to conduct the attack at dynamically changing scenes. This includes varying lighting conditions, varying distances, e.g., between a sign and an approaching vehicle, and sometimes in the presence of motion blur induced by the vehicle's movement. 

\vspace{0.5 em} \noindent\textbf{Cost-effective deployment.} The adversary is clearly interested in minimizing the amount of time and the cost required to mount such attacks. Namely, the generation of physical adversarial perturbations and their deployment should require minimal time and resources.

\subsection{Approach}

A strawman solution to create inconspicuous adversarial examples would be to rely on infrared projectors (e.g., Barco FS70-W6). Such projectors are widely used in military applications and are notorious for their ability to introduce precise pixel-wise projections.  
Infrared projectors are, unfortunately,  costly (in the order of tens of thousands of USD).  

In this work, we opt for a more efficient alternative to infrared projectors. Namely, we explore using a transparent film on which we print the perturbation combined with the mask using an off-the-shelf printer. Our primary intuition is to discreetly place the film in front of an infrared light source, allowing it to project the perturbation onto our target object (cf. {\Cref{fig:system_model}}). Notice that this process precisely mimics the projection of small squares onto the target image. Our setup consists of a compact device that integrates both an infrared lamp and transparent film {(approx. $10\mathrm{cm}\times 10\mathrm{cm} \times 20\mathrm{cm}$)} which can be \emph{easily concealed behind bushes or junction boxes}. This is considerably more stealthy than setups used in prior works, which, for example, require bulky video projectors \cite{lovisottoSLAPImprovingPhysical} or infrared lasers \cite{satoInvisibleReflectionsLeveraging2024}, that can also pose a safety hazard due to the use of lasers. 

In this setting, several challenges arise to ensure that our design is scene- and system-agnostic. Namely, while the reliance on the infrared domain presents opportunities to the adversary, it effectively limits the available color space that can be utilized when creating perturbations (and increases the difficulty of successfully generating adversarial examples). Moreover, one needs to ensure that adversarial examples are efficiently created to be effective in real-world deployments practically and to adjust to environmental changes quickly; for instance, we need to cater to the fact that the CMOS camera is constantly moving (and is not fixed when compared to other use cases) and, thus, lighting and position would also vary as the vehicle approaches the traffic sign.

\section{Design} \label{sec:design}
We now present our methodology for generating efficient and workable adversarial perturbations in the infrared domain. 

\subsection{Modeling Infrared Perturbations} \label{sec:modeling_perturbations}
Unlike traditional perturbation attacks, our infrared perturbations must be created while paying special attention to the fact that there might be multiple light sources involved (i.e., an ambient and an infrared light source); this complicates the modeling process significantly. We summarize the various notations used in this paper in~\Cref{tab:mp_parameters}.

\begin{table}[t]
\footnotesize
	\caption{Overview of \manypixel parameters.}
	\centering 
	\begin{tabular}{ll}
     \toprule
    Parameter &  Description \\
    \midrule
   $w$ & Width of the input image. \\
   $h$ &Height of the input image. \\
   $k$ & Maximum number of \manypixel used in the perturbation. \\
   $l$ & Side length of an \manypixel in pixel.\\
   $\mathcal{I}$  & Set of \manypixel positions in the reduced coordinate space. \\
   $\mathcal{\tilde{I}}$ & Set of pixel positions in the pixel coordinate space. \\
    $\mathcal{M}$ & Mask to locate the target object. \\
   $\mathcal{P}$ & Projection mask consisting of the mask and \manypixel. \\
   \bottomrule
	\end{tabular}
 \label{tab:mp_parameters}
\end{table}

\begin{figure}[tb]
     \centering
  \subfloat[\manypixel positions.\label{fig:mask_positions}]{\makebox[1.1\width]{\centering\includegraphics[width=.25\linewidth]{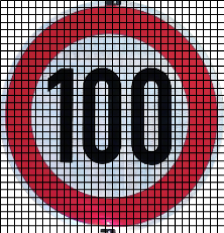}}}
  \hspace{0.1 em}
    \subfloat[Perturbation mask.\label{fig:sample_perturbation}]{\makebox[1.1\width]{\centering\includegraphics[width=.26\linewidth]{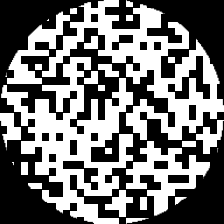}}}
    \hspace{0.1 em}
    \subfloat[Infrared perturbation.\label{fig:sample_transformation}]{\makebox[1.25\width]{\centering\includegraphics[width=.26\linewidth]{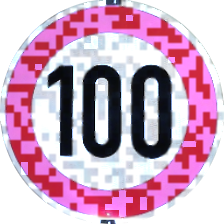}}}
  \caption{Definition of pixel positions ($32\times32$) and a concrete perturbation for $l=7$ and an image of $w=h=224$.}
\label{fig:patch_modelling}
\end{figure}

\vspace{0.25 em}\noindent \textbf{Shape.}
Due to the imperfect nature of the perturbation process, we opted to move away from pixel-wise perturbations to so-called \emph{manypixel} (\textsf{MP}), a grouping of several neighboring pixels. 
For simplicity and without loss of generality, we assume that a \manypixel can be approximated by a square whose side length $l$ divides both the height and width of the input image, i.e., $l|h \wedge l|w$. This effectively reduces the pixel coordinate space from $w \times h$ pixels to a reduced coordinate space of  $\nicefrac{h}{l} \times \nicefrac{w}{l}$ \manypixel for our subsequent optimizations, as seen in \Cref{fig:mask_positions}. This matches our real-world experiments in \Cref{sec:real_world_exp}, where a square \manypixel is output by projecting small pixel perturbations from a transparent film onto the target.

\vspace{0.25 em}\noindent \textbf{Location.}
We define the location of our adversarial perturbation by a set of \manypixel positions $\mathcal{I}$. For instance, when the \manypixel corresponds to a square, 
$\mathcal{I} \subseteq [0, \nicefrac{w}{l}] \times [0, \nicefrac{h}{l}]$. The amount of \manypixel is denoted by $k$, i.e., $|\mathcal{I}|=k$.
A mask $\mathcal{M}$ is used to locate the target object, e.g., the shape of a traffic sign and the region of the adversarial perturbation. It might happen that an \manypixel is drawn directly on the mask. To enable a partial drawing of the \manypixel along the contour of the mask, we define a transformation from the reduced coordinate space used for the \manypixel back to the pixel coordinate space of the original input image. In the particular case of a square, this is achieved by adding the positions of all $l^2$ pixels within a single \manypixel to our position set. We define the transformation $\varphi$ as follows:
\begin{equation}
    \varphi(x,y) := [x, x+l] \times [y, y+l]
\end{equation}
and construct the transformed coordinates $\mathcal{\tilde{I}}$ as
\begin{equation}
    \mathcal{\tilde{I}} := \bigcup_{(x,y) \in \mathcal{I}}\varphi(x,y)
\end{equation}
Using the aforementioned transformation, the function $\mathtt{ModelPerturbation(\mathcal{I})}$ returns the projection mask $\mathcal{P}$ that corresponds to the intersection of the \manypixel locations in pixel space and the mask, i.e., $\mathcal{P} = \mathtt{ModelPerturbation(\mathcal{I})} = \mathcal{\tilde{I}} \cap \mathcal{M}$ as shown in \Cref{fig:sample_perturbation}.

\vspace{0.25 em}\noindent \textbf{Perturbation color.}
We need to model the impact of the infrared light source on an object that is already lit by ambient light. Our perturbation masks the infrared light source, making the covered area appear as ambient-lit. As a result, the area outside the perturbation exhibits a color shift while the perturbed area remains unaffected. Unlike prior work that relies on a single visible light source \cite{zhongShadowsCanBe2022}, we cannot assume that perturbations can be modeled simply as brightness reductions.

\begin{figure}[tb]
\centering\includegraphics[width=0.7\linewidth]{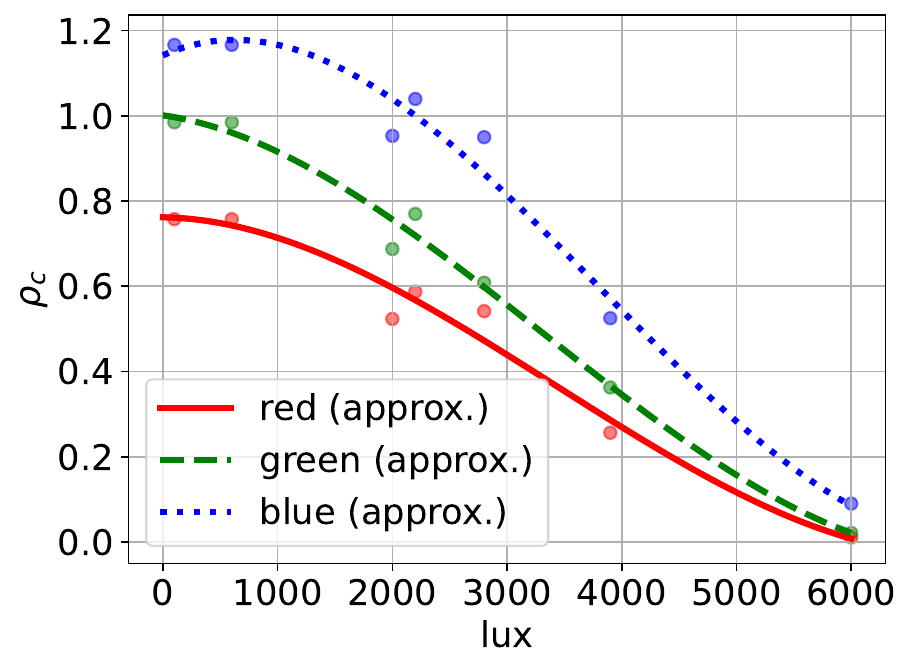}
  \caption{Estimated channel-specific scaling factors $\rho_c$ for various ambient lighting intensities. 
  }
\label{fig:est_fun}
\end{figure}

\begin{figure}[tb]
     \centering
    \centering\includegraphics[width=0.75\linewidth]{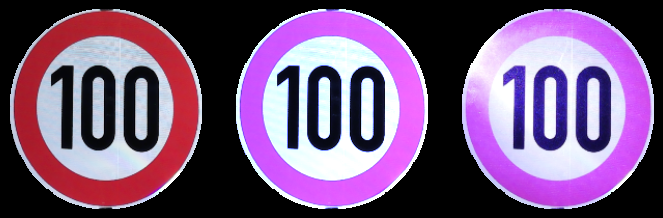}
  \caption{Comparison of a real-world infrared light source (right), a simulated infrared light source (center) for a traffic sign (left) from the GTSRB dataset.
  }
\label{fig:comparison_simulation}
\end{figure}

Note that, in ideal conditions, cameras adjust their exposure and white balance to obtain similarly bright images, albeit being taken under different lighting conditions. However, many real-world traffic sign datasets contain overexposed or underexposed images due to complex lighting scenarios, which obscures the impact of an infrared light source. Before introducing a brightness-dependent infrared transformation, we normalize the data to ensure that we obtain similarly exposed images. This can be modeled using the three-dimensional CIELAB color space~\cite{iso_cie_11664_4_2019}, which covers the entire gamut of human color perception. More concretely, the lightness channel $L$ correlates with the perceptual lightness, and we assume that an exposure adjustment only changes this channel. 
The $A$ and $B$ channels model the four unique colors of human perception, i.e., red, green, blue, and yellow, and remain unchanged. 
We define the transformation from RGB to CIELAB space as follows: 
\begin{align} \label{ycrcb_trans}
   \mathbf{LAB}: [0,255]^{3} &\rightarrow [0, 100] \times [-128, 127]^2\\
    \mathbf{LAB}(x) &= [ \mathbf{L}_x\;\mathbf{A}_x\;\mathbf{B}_x ]
\end{align}
The resulting normalization adjusts the average lightness value $\mathbf{\bar{L}}_x$ of an image $x$ to the average lightness value of the data set $\bar{L}_\mathrm{data}$ as follows:
\begin{align*}
    \mathtt{Normalize}(x) = \mathbf{LAB}^{-1}\Big(\Big[\frac{\mathbf{L}_x{\bar{\mathbf{L}}}_x}{\bar{L}_\mathrm{data}} \;\mathbf{A}_x\;\mathbf{B}_x\Big]^T\Big)
\end{align*}
Considering the fixed position and power of an infrared light source, the largest impact is observed on the red channel, with less pronounced effects on the green and blue channels. Due to the fixed power of the infrared light source, its effect is attenuated as the ambient lighting increases. To model the color channel of an infrared image ($\mathbf{IR}_c$) based on these effects, we take the visual color channel ($\mathbf{VIS}_c$) and apply a channel and ambient lighting-specific scaling ($\rho_c$) of the red color channel ($\mathbf{VIS}_r$). This relationship is captured in the $\mathbf{IR}$ transformation as follows:
\begin{align}
\mathbf{IR}&: [0, 255]^3 \rightarrow [0, 255]^3 \\
    \mathbf{IR} &= [ \mathbf{IR}_r\;\mathbf{IR}_g\;\mathbf{IR}_b ]\\
        \mathbf{IR}_c &= {\mathbf{VIS}_c + \mathbf{VIS}_r * \rho_c}\label{eq:ir_trans}
\end{align}
To estimate the scaling parameter $\rho_c$ for various ambient lighting intensities, we rely on empirical measurements. Here, we conducted experiments in a range of $100-6000$ lux on the surface of a traffic sign. Based on the resulting pairs of images with only the ambient lighting and an additional infrared light source, we estimated the scaling parameter as follows:
\begin{equation} 
        \rho_c = \frac{\mathbf{IR}_c - \mathbf{VIS}_c}{\mathbf{VIS}_r}
\end{equation}
We used these data points to fit channel-specific functions, as shown in \Cref{fig:est_fun} to perform the transformation digitally.
An example of a successful transformation is shown in \Cref{fig:comparison_simulation}.
In practical scenarios, we find that the scaling parameter $\rho_c$ depends on various factors, most prominently ambient lightning.
To ensure the robust generation of physical adversarial examples, we account for some variation of this parameter with EOT \cite{athalyeSynthesizingRobustAdversarial} (cf. \Cref{sec:real_world_pert}).

We define a function $\mathtt{ApplyIR}$ that takes an input image $x$ and the projection mask $\mathcal{P}$ and applies the infrared transformation (cf. \Cref{eq:ir_trans}) only to the parts of the image that are \emph{not} covered by an \manypixel as a means to prevent the infrared light from reaching the surface of the traffic sign. This is achieved using the Hadamard product as follows:
\begin{equation}
    x'= \mathtt{ApplyIR}(x, \mathcal{P}) = x \odot \mathcal{P} + \mathbf{IR}(x) \odot (1 - \mathcal{P})
\end{equation}
and is shown for an example in \Cref{fig:sample_transformation}.

\subsection{Optimization for Two-Stage Architectures} \label{optimization_strategy}

We start by describing our optimization strategy for the challenging two-stage architecture. In \Cref{single-stage-optim}, we also outline the optimization strategy for the single-stage architecture. 
For image classification, let~$f_\theta\colon \RR^d\to \Delta^n$ denote a DNN model, parameterized by $\theta$, assigning $d$-dimensional inputs to~$n$ classes, where~$\Delta^n$ is the probability simplex of~$n$ classes, and let~$\Classifier\colon \RR^d\to [n]$ refer to the associated classifier defined as $\Classifier(x) := \argmax_{i \in [n]} f_i(x)$. The dimension is equivalent to the number of pixels, i.e., $d=h\times w \times c$, with width $w$, height $h$, and number of color channels $c$. Given a genuine input~$x\in \RR^d$ predicted as~$C(x) = s$ (source class), desired target class $t$, and an adversarial perturbation $\delta \in \RR^d$, $x'=x+\delta$ is considered an \emph{adversarial example} of~$x$ if the following criterion is fulfilled: 
\begin{equation}
    \label{eq:misclassifications}
    \mathcal{A}(x'):=
    \begin{cases}
        C(x') \neq s & \text{(untargeted attack),} \\
        C(x') = t    & \text{(targeted attack).}
    \end{cases}
\end{equation}
The objective of the adversary is then expressed with the following margin loss function~\cite{carliniEvaluatingRobustnessNeural2017}: 
\begin{equation}
    \label{eq:adversarial:objective}
    \loss_{adv}(x) :=
    \begin{cases}
        f_s(x) - \max\limits_{i\neq s}f_i(x)    & \text{(untargeted attack),}\\
        \max\limits_{i\neq t} f_i(x) - f_{t}(x) & \text{(targeted attack).}    
    \end{cases}
\end{equation}
For an adversarial example~$x'$ to be successful, we require that there is a perturbation $\mathcal{P}$ that satisfies~$\loss_{adv} < 0$ to achieve an (un)targeted misclassification. The optimization problem is defined as follows
\begin{eqnarray}
	&\min \loss_{adv}(  \mathtt{ApplyIR}(\mathtt{Normalize}(x_{\mathrm{input}}), \mathcal{P})) \label{eq:adversarial_objective}
\\
\quad\text{s.t.}\quad
&\mathcal{P} = \mathtt{ModelPerturbation(\mathcal{I})}
\end{eqnarray}
to find an optimal set of \manypixel positions $\mathcal{I}$ for a given input image $x_{\mathrm{input}}$.
Based on \Cref{eq:misclassifications} and the subset of a given dataset $\mathcal{S}$, we define the attack success rate ($\mathsf{ASR}$) for the digital and physical experiments as follows:
\begin{equation}\label{eq:asr}
    \mathsf{ASR} := \nicefrac{1}{|\mathcal{S}|}\textstyle\sum_{x \in \mathcal{S}}^{} \mathbf{1}(\mathcal{A}(x'))
\end{equation}
with output $x'$ of $\mathtt{ApplyIR}(x_{\mathrm{input}})$, adversarial criterion $\mathcal{A}$, and the indicator function defined as $\mathbf{1}(x)= 1$ if $x$ is true and $\mathbf{1}(x)= 0$, otherwise.

In contrast to existing works that need to perform complex modeling of adversarial shapes and locations~\cite{zhuFoolingThermalInfrared2021, zhuInfraredInvisibleClothing2022, weiHOTCOLDBlockFooling2023, weiPhysicallyAdversarialInfrared2023,satoInvisibleReflectionsLeveraging2024} to remain inconspicuous and/or easily manufacturable, due to our large perturbation space, we can resort to the well-known random search algorithm \cite{rastrigin1963convergence} for the derivative-free optimization of our problem. Since, in our use case, it may not be possible to simply extract a model from the automotive system for more powerful attacks, we opt to investigate a more realistic threat model, treating our system as a black box, which is generally considered more applicable to real-world deployments.

\begin{figure}[tb]
     \centering
       \subfloat[It. 0, $k=64$]{\centering\includegraphics[height=2.2cm]{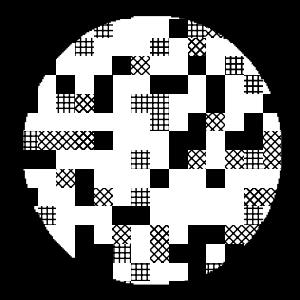}}\qquad
       \subfloat[It.  20, $k=52$]{\centering\includegraphics[height=2.2cm]{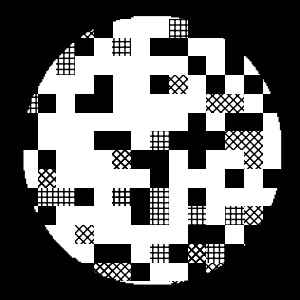}}\qquad
       \subfloat[It.  160, $k=13$]{\centering\includegraphics[height=2.2cm]{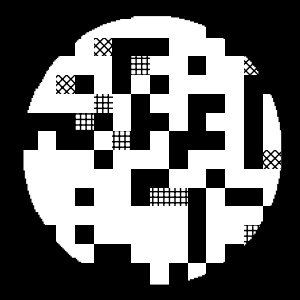}}   
  \caption{Selected iterations from our approach. Crossed hatches indicate the addition of a \manypixel from a previous iteration, while diagonally crossed hatches indicate a removal of a \textsf{MP} from a previous iteration.}
\label{fig:exp_convergence}
\end{figure}

As mentioned earlier, we approximate a \manypixel with a square (see \Cref{sec:modeling_perturbations} for the justification) and proceed to find a (locally) optimal set of \manypixel positions that cause an (un)targeted misclassification. To this end, we randomly perturb a total of $k$ different \manypixel, i.e., $\mathcal{I} \leftarrow \mathcal{U}(0, \nicefrac{h}{l}) \times \mathcal{U}(0, \nicefrac{w}{l})$, throughout a maximum of $Q$ queries to the classifier, {with $\mathcal{U}$ denoting the uniform distribution}. Naively perturbing the \manypixel until we converge to a successful adversarial example is intractable, especially in the targeted attack setting. Instead, we only update the current best set of \manypixels{} if the new set of \manypixels{} results in a lower loss. To improve convergence, we exponentially decrease the number of changed \manypixel based on the current iteration. Initially, we change the largest number of \manypixel to significantly reduce the loss (cf. \Cref{eq:adversarial:objective}), while subsequent queries with fewer changed \manypixel are intended to refine the loss. The schedule for deriving the number of perturbed \manypixel for a given iteration $i$, query budget $Q$, and maximum number of perturbed \manypixel $k$ is defined as follows:
$k(i) = \Big\lceil\frac{k}{2}e^{\frac{\ln\frac{2}{k}}{Q} \cdot i}\Big\rceil$. 

This schedule is used to randomly draw $k(i)$ new \manypixel from the image. Before adding them to the set of indices $\mathcal{I}$, we randomly remove $k(i)$ \manypixel from this set.
Once we obtain a negative loss, the goal of (un)targeted misclassification is achieved. 
The overall algorithm for generating infrared perturbations is shown in Algorithm~1. Given an initial image, this algorithm outputs the perturbation mask $\mathcal{P}$ and the resulting infrared adversarial image $x'$. The process of optimizing the \textsf{MP}s is shown in \Cref{fig:exp_convergence}. 

\SetKwComment{Comment}{/* }{ */}

\begin{algorithm}[tb]
\footnotesize
\DontPrintSemicolon
\caption{Generating infrared perturbations}
\label{alg:two}
\KwData{input $x$, loss $\loss$, max query $Q$, number of \textsf{MP} $k$, \textsf{MP} size $l$ 
}
\KwResult{candidate for minimizing $\mathcal{L}$}
\Comment{Initialize first candidate}
$\mathcal{I} \gets \mathcal{U}(0, \nicefrac{h}{l}) \times \mathcal{U}(0, \nicefrac{w}{l}); |\mathcal{I}|=k$\; 
$\mathcal{P} = \mathtt{ModelPerturbation(\mathcal{I})}, x'= \mathtt{ApplyIR}(x, \mathcal{P})$\;
$\loss^* \gets \loss(x'), i \gets 0$\;
\While{$i < Q$ }{
\Comment{Update positions of perturbation}
  $\mathcal{I}' \gets \mathcal{U}(0, \nicefrac{h}{l}) \times \mathcal{U}(0, \nicefrac{w}{l}); |\mathcal{I}'|=k(i)$\; 
  $\mathcal{I}'' \gets$ randomly select $k(i)$ indices from $\mathcal{I}$\; 
  $\mathcal{I}' \gets (\mathcal{I} \setminus \mathcal{I}'') \cup \mathcal{I}'$\; 

$\mathcal{P}' = \mathtt{ModelPerturbation(\mathcal{I'})}, x'= \mathtt{ApplyIR}(x, \mathcal{P}')$\;
  \Comment{Upon improvement, update solution}
  
  \If{$\loss(x') < \loss^*$}{
      $\loss^* \gets \loss(x'), \mathcal{I} \gets \mathcal{I}'$\;
  }
  \Comment{Negative loss: success}
  \If{$\loss^* < 0$}{
      break\; 
  }
  $i \gets i + 1$\;
}
$\mathcal{P} = \mathtt{ModelPerturbation(\mathcal{I})}, x'= \mathtt{ApplyIR}(x, \mathcal{P})$\;
\Return{$\mathcal{P}, x'$}
\end{algorithm}

\begin{figure}[tb]
     \centering
       \includegraphics[width=0.75\linewidth]{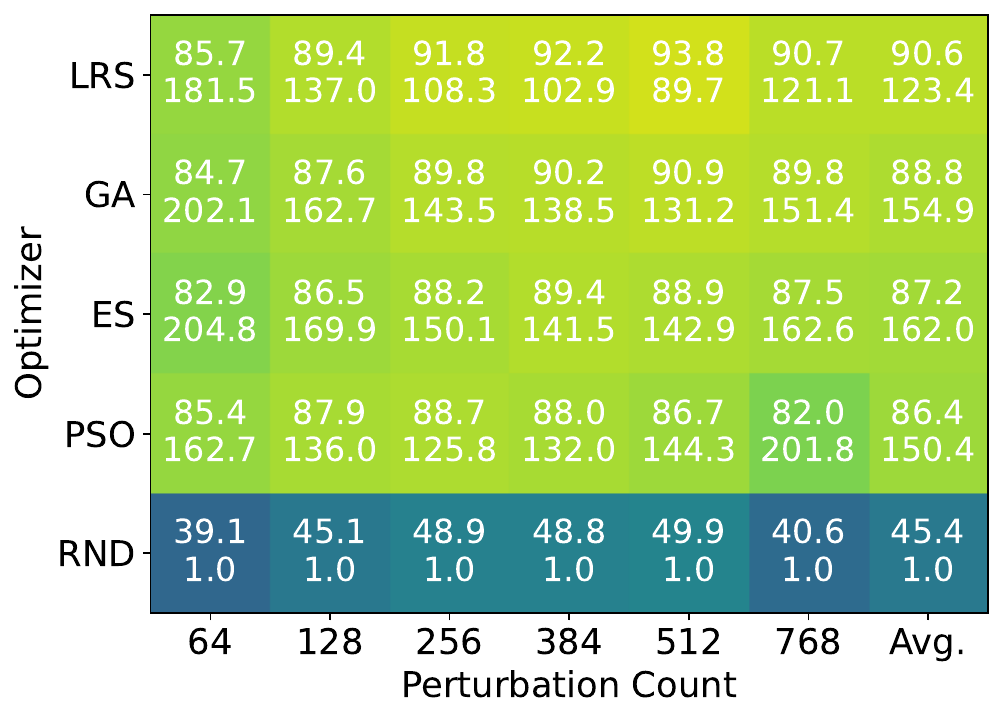}\\
  \caption{Optimization results for a two-stage architecture on the GTSRB dataset. Each cell contains the \asr and query count. The last column includes the averages.}
\label{fig:results_optimization_selection}
\end{figure}

\vspace{0.25 em}\noindent \textbf{Comparison of Optimization Strategies.} 
To confirm the superiority of our optimization, we now compare the \asr and average queries achieved by various popular optimization strategies, i.e., local random search (LRS), particle swarm optimization (PSO), genetic algorithms (GA), and evolution strategies (ES) for an ambient light setting of $10$ lux and ablate the number of \manypixel $k$ of size $l=1$ in an untargeted attack setting. We compare our results against a baseline consisting of a naive random strategy (RND) that randomly places up to $k$ \manypixel.  Our results are depicted in \Cref{fig:results_optimization_selection} for GTSRB~\cite{stallkampManVsComputer2012} with $25$ samples for each of the $43$ classes.

We observe that \emph{local random search} results in the highest \asr over all perturbation counts $k$ with an average of $90.6\%$ (i.e., twice as high compared to a random positioning of perturbations) and with as few as $123.4$ queries. 

\subsection{Optimization for Single-Stage Architectures}
\label{single-stage-optim}
To detect objects, let~$f_\theta\colon \RR^d\to \{ (\texttt{bbox}, \Delta^n) \}^m$ denote a DNN model, parameterized by $\theta$, assigning $d$-dimensional inputs to~$m$ bounding boxes \texttt{bbox}, each with a probability simplex~$\Delta^n$ of~$n$ classes. 
Without loss of generality, we focus on an image with one bounding box. We let~$\Detector\colon \RR^d\to \{ (\texttt{bbox}, [n]) \}$ refer to the associated detector which is defined as $\Detector:=\{(\texttt{bbox}, \argmax_{i \in [n]} f_i(x)) \}$ for each bounding box for which the maximum probability is above the detection threshold $\tau$, i.e., $\max_{i \in [n]} f_i(x) > \tau$.

We consider a genuine input~$x\in \RR^d$ predicted as~$D(x) = \{(\texttt{bbox}, s) \}$ (source class) and define the adversarial criterion and objective for a hide attack, i.e., {not recognizing the bounding box of the source class}, as follows: 
\begin{align}
    \label{eq:misclassifications_od}
    \mathcal{A}(x')&:=\Big\{\{(\texttt{bbox}, s)\} \notin D(x')\\
    \loss_{adv}(x)&:=\Big\{\max_{i \in [n]} f_i(x) - \tau     
\end{align}
For an adversarial example~$x'$ to be successful, we require that there is a perturbation $\mathcal{P}$ that satisfies~$\loss_{adv} < 0$ to achieve a hide attack. The optimization problem is then identical to the previous case of image classification (cf. Equation \eqref{eq:adversarial_objective}). We use Equation~\eqref{eq:asr} to compute the attack success rate.

\begin{figure}[tb]
     \centering
       \includegraphics[width=0.75\linewidth]{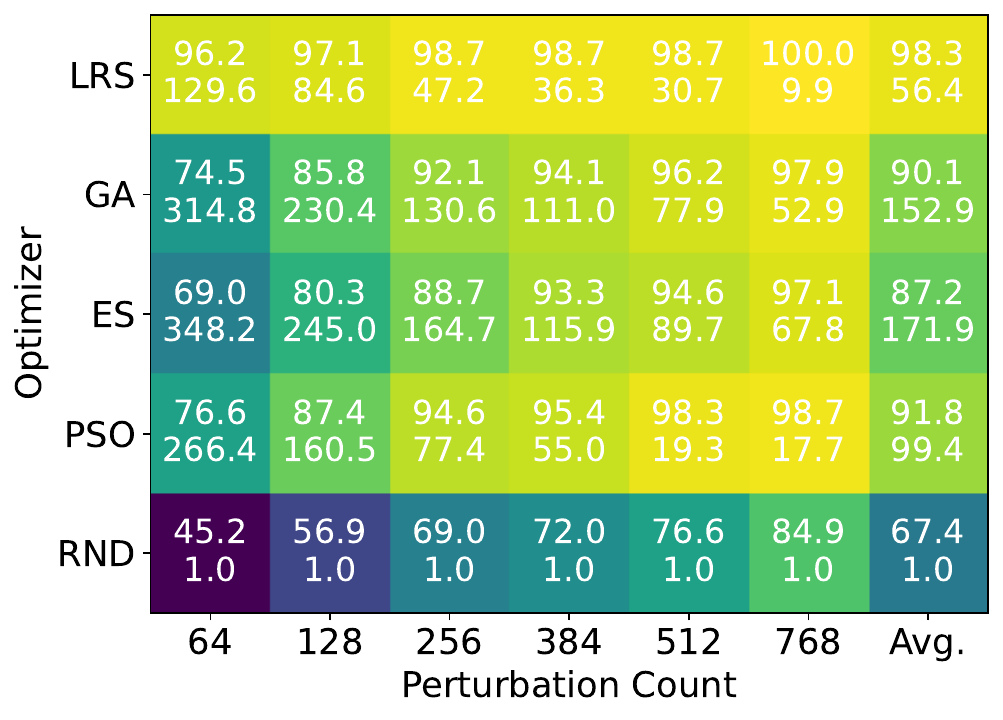}
  \caption{Optimization results for a single-stage architecture on the Mapillary dataset. Each cell contains the \asr and query count. The last column includes the averages.}
\label{fig:results_optimization_mapillary}
\end{figure}

To evaluate the impact of the optimizer selection, we evaluated the \asr and average consumed queries achieved with various popular optimization strategies on the YOLOv8 model \cite{glennjocherandayushchaurasiaandjingqiuUltralyticsYOLOv82023} trained on the Mapillary\cite{10.1007/978-3-030-58592-1_5} dataset. Our results are depicted in \Cref{fig:results_optimization_mapillary}. Here, we use $25$ samples for each of the $9$ classes, i.e., speed limits and stop, with a total of $225$ images (cf. \Cref{single-stage-eval}). In line with our previous results, the local random search optimization algorithm performs best, with an average attack success rate of $98.3\%$ and an average of $56.4$ consumed queries, outperforming the second-best optimization procedure, particle swarm optimization, with an average \asr of $91.8\%$ and an average of $99.4$ consumed queries. 

\subsection{Validating our Infrared Model} \label{sec:validation_transform}

To validate our model, we created an (open-source) infrared traffic sign dataset. Our dataset comprises images of traffic signs with varying levels of ambient lighting, both with and without an additional infrared light source.  We include additional details about our dataset in Appendix~\ref{sec:dataset_ir}. We compare the success of our approach on (1) the real infrared images and (2) the emulated infrared light stemming from our digital transformation in~\Cref{sec:modeling_perturbations}.
We conducted our experiments with $l=14$, in line with~\Cref{sec:real_world_exp}.

As shown in~\Cref{tab:validation}, our results on the real-world dataset show an across-the-board \asr of $100\%$ with an averaged consumed queries as low as $21.7$ for $k=96$. For our simulated infrared light source, we observe the highest \asr at $k>128$ with $94\%$, while the lowest observed rate ranks at $88\%$. Our results, however, confirm that our infrared transformation provides a \emph{tight worst-case emulation of the real-world}. This also means that we expect our approach to yield, on average, better success rates in the real world when compared to the digital world.

\subsection{Real-World Physical Perturbations} \label{sec:real_world_pert}
To ensure robustness of adversarial examples under real-world conditions, we rely on expectation over transformation (EOT) \cite{athalyeSynthesizingRobustAdversarial} that finds a perturbation over the expected value of all transformed inputs over the set of transformations $\Omega$:
\begin{align*}
	&\min \mathbb{E}_{\omega\sim \Omega}[\loss_{adv}(\omega( \mathtt{ApplyIR}(x_{\mathrm{input}}, \mathcal{P})))]\\
\quad\text{s.t.}\quad
&\mathcal{P} = \mathtt{ModelPerturbation(\mathcal{I})}
\end{align*}
We model only reasonable effects with justified value ranges, as overly complex transformations result in difficult convergence towards a suitable adversarial example. More concretely, $\Omega$ includes transformations for the following effects~\cite{lovisottoSLAPImprovingPhysical}: 

\vspace{0.25 em} \noindent \textbf{Perspective.}
Traffic signs are typically placed on the right side of a street (in countries with right-hand traffic) at a typical height of 2m in Europe and 5-7 ft in the US. We assume that the camera is placed at an average height of a European vehicle of 1.5m. As a result, we consider both an $x$-axis and $y$-axis perspective transformation of $\pm 35 \deg$. 

\vspace{0.125 em} \noindent \textbf{Distance.}
As a vehicle approaches a traffic sign, the initially small sign gets larger over time in the captured images. This results in an initial upsampling, followed by a subsequent downsampling once the sign is too large for the network to process. For the lower bound on distance, we determine the minimum sign size for which the DNN can still correctly classify a given sign. As a result, we determine the minimum size to be $18\times18$ pixel.

\vspace{0.125 em} \noindent \textbf{Rotation.}
Traffic signs are typically mounted straight, i.e., the horizontal sign axis is perpendicular to the street. Due to imperfect mounting, we tolerate rotations of $\pm 6 \degree$.

\vspace{0.125 em} \noindent \textbf{Brightness.}
To account for the slight over-/underexposure of a camera, we also utilize the LAB color space to model a brightness change (cf. \Cref{ycrcb_trans}). We consider a value range of $\pm 20\%$ in the lightness $L$ channel of the image.

\vspace{0.125 em} \noindent \textbf{Backgrounds.}
In single-stage pipelines, the background can significantly influence the model's output. To ensure robustness across various settings, we place the sign against a variety of backgrounds.

\vspace{0.125 em} \noindent \textbf{Alignment and Motion Blur.}
Due to the required alignment of the perturbation onto the traffic sign, we consider a shift in the $x$- and $y$-axis of $\pm {5} $pixel. We also introduce motion blur to mimic blur on frames of a moving camera.

To implement our setup in~\Cref{fig:system_model}, the film must be carefully aligned with the light source, which can be efficiently done using an infrared camera as a viewfinder. We also added a 3D-printed magnetic frame to prevent film bending and projection distortions.

\begin{table}[tb]
\footnotesize
	\caption{Comparison of \asr and average consumed queries $Q$ on real-world and simulated infrared perturbations based on our model in~\Cref{sec:modeling_perturbations}.}
	\centering 
	$\begin{array}{l|cc|cr}
     \toprule
    k &  \multicolumn{2}{c|}{\textbf{Real-World}}  &  \multicolumn{2}{c}{\textbf{Simulated}}   \\ 
   (\# \manypixel) & \asr & Q & \asr & Q \\\midrule
   16 & 100.0 & 36.94 & 88.0 & 164.38   \\ 
   32 & 100.0 & 41.28 & 88.0 & 144.84   \\ 
   64 & 100.0 & 21.70 & 90.0 & 117.44   \\ 
   96 & 100.0 & 11.20 & 90.0 & 125.38   \\ 
   128 & 100.0 & 41.98 & 94.0 & 77.32   \\ 
   192 & 100.0 & 17.66 & 94.0 & 97.38   \\ 
   \bottomrule
	\end{array}$
 \label{tab:validation}
\end{table}

\section{Experiments on Two-Stage Architectures} \label{sec:experiments}
In this section, we empirically evaluate our approach for traffic sign recognition in the digital and physical domains. 

\vspace{0.125 em} \noindent \textbf{Datasets and Models:} We conducted our experiments on established datasets for traffic sign recognition for two-stage architectures: we rely on GTSRB ~\cite{stallkampManVsComputer2012} for German traffic signs and LISA~\cite{mogelmoseVisionbasedTrafficSign2012} for American traffic signs. 
For the underlying model architectures, we use a simple CNN~\cite{yadav} for GTSRB and LISA-CNN (taken from the cleverhans library \cite{goodfellowCleverhansV01Adversarial2016}), which is in line with previous works in this field \cite{eykholtRobustPhysicalWorldAttacks2018, zhongShadowsCanBe2022, lovisottoSLAPImprovingPhysical}. 
In our normalized test sets, we report a clean accuracy (\textsf{CA}) of {$98.76\%$ and $99.63\%$} for GTSRB and LISA, respectively.

\subsection{Targeted Attacks in the Digital Domain}
\label{targeted_eval}
\begin{figure}[tb]
     \centering
\subfloat[Scenario 1 (brake).\label{fig:scenario_1}]{\centering\includegraphics[width=.47\linewidth]{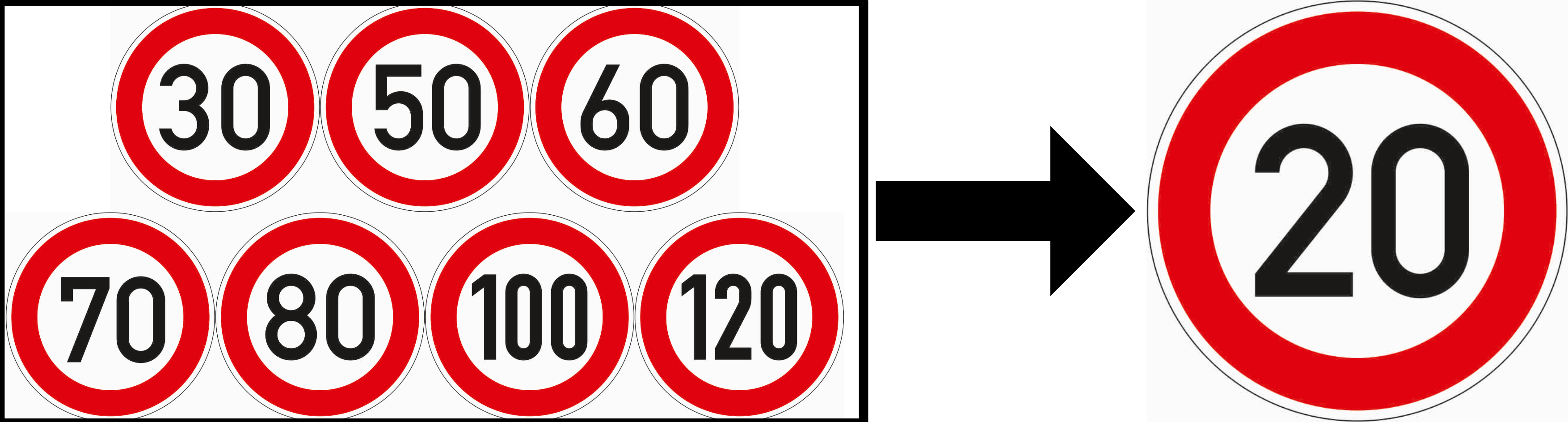}\hspace{0.45em}\rule{1.5pt}{1cm}\hspace{0.45em}\includegraphics[width=.47\linewidth]{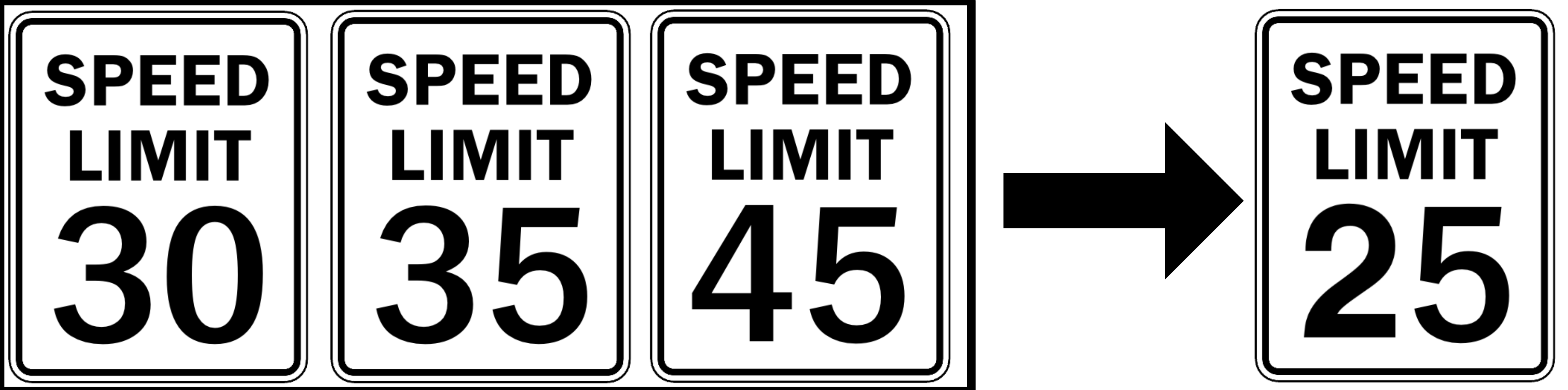}}\\
            \subfloat[Scenario 2 (acceleration).\label{fig:scenario_2}]{\includegraphics[width=.47\linewidth]{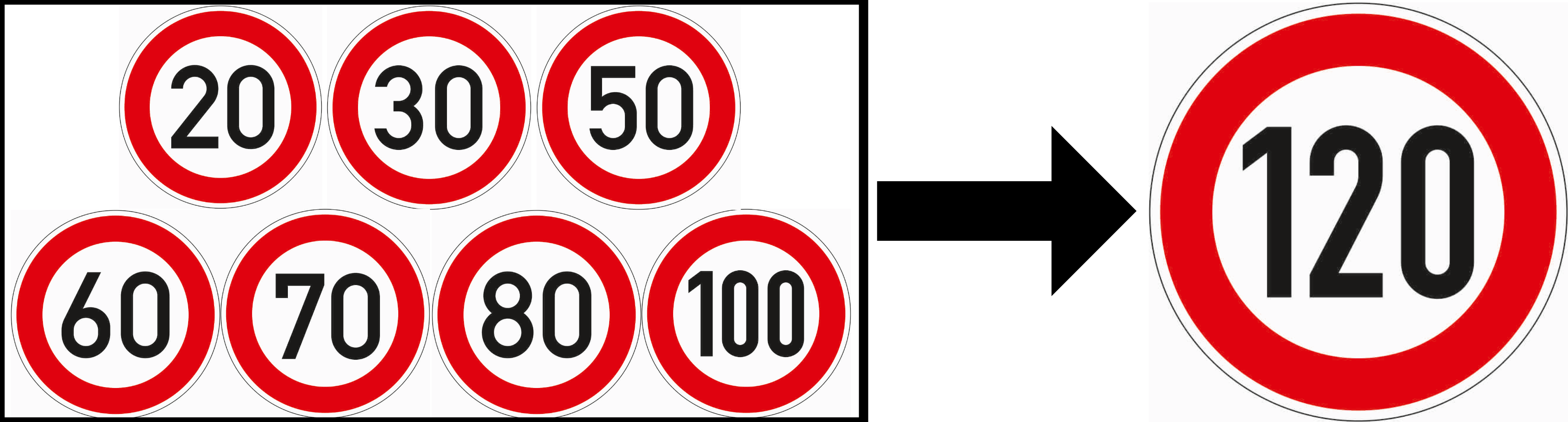}\hspace{0.45em}\rule{1.5pt}{1cm}\hspace{0.45em}\includegraphics[width=.47\linewidth]{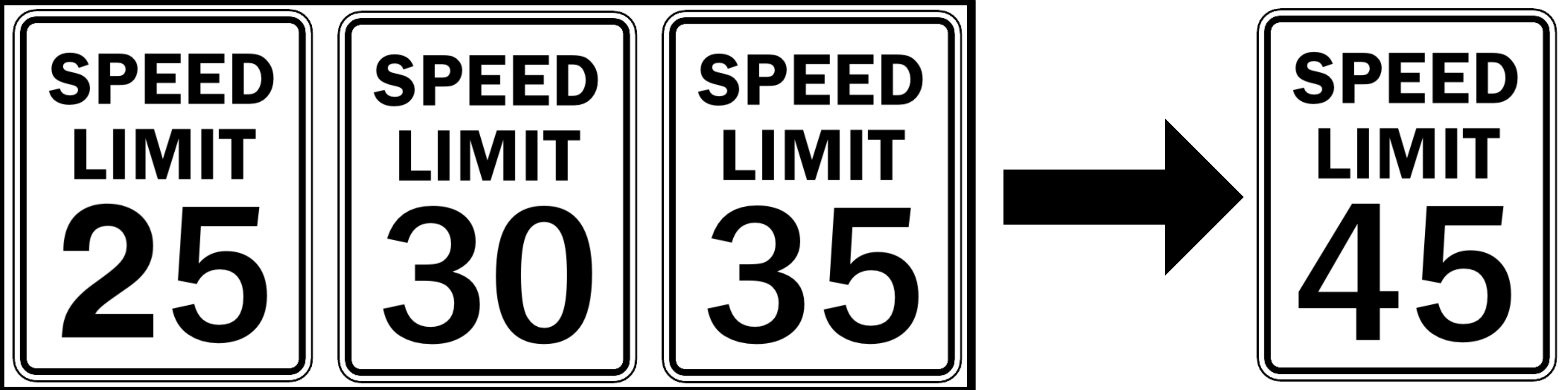}}\\
              \subfloat[Scenario 3 (ignore stop).\label{fig:scenario_3a}]
             {\centering\includegraphics[height=0.95cm]{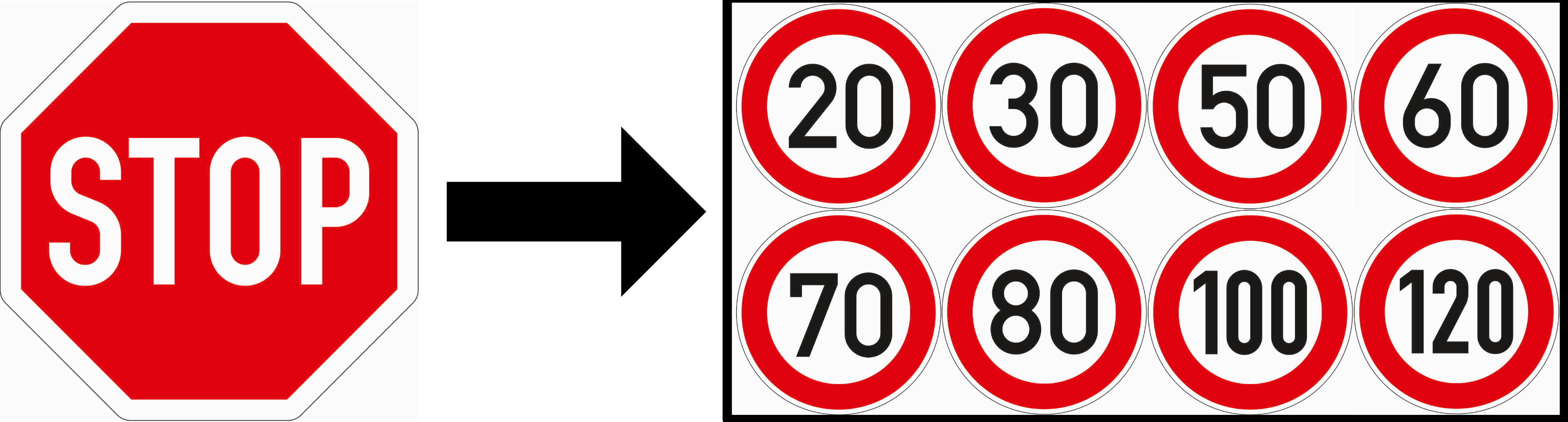}\hspace{0.2em}\rule{1.5pt}{1cm}\hspace{0.2em}\includegraphics[height=0.95cm]{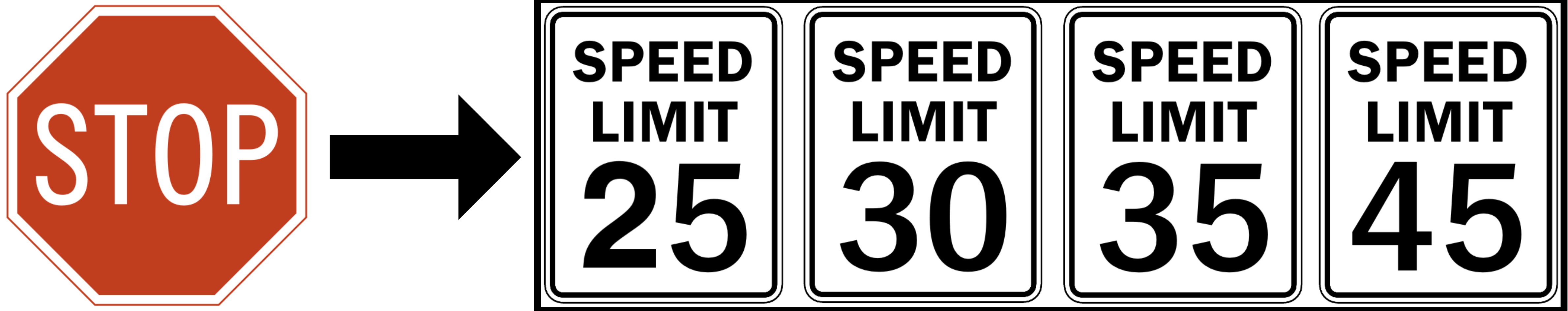}}%
  \caption{Overview of the targeted class flips in our scenarios with European traffic signs on the left and North American traffic signs on the right.}
\label{fig:scenarios}
\end{figure}

Our evaluation in the digital domain emulates physical attacks in the real world using the GTSRB and LISA datasets. We start by evaluating our approach in the more challenging targeted attack scenario, where an adversary seeks to ensure that the prediction only flips to a \emph{specific} class. In \Cref{untargeted_eval}, we also discuss the effectiveness of our approach in the untargeted setting. For a meaningful evaluation of a (semi\nobreakdash-)autonomous vehicle, we define the following three driving scenarios that result in prominent safety hazards, especially when triggered in a \emph{targeted} manner. Concretely, they result in a reduction of speed, i.e., braking, acceleration, or the ignoring of a stop sign, because a speed sign is recognized. We show them in \Cref{fig:scenarios}. 

\vspace{0.25 em}\noindent \textbf{Scenario 1 (Brake).} We use all speed sign classes (except the lowest one) as source images and generate an adversarial example for each that classifies it as the lowest speed.

\vspace{0.25 em}\noindent \textbf{Scenario 2 (Acceleration).} We use all speed sign classes (except the highest) as source images and generate an adversarial example for each that classifies it as the highest speed.

\vspace{0.25 em}\noindent \textbf{Scenario 3 (Ignore stop).} We use the stop sign class as source images and generate adversarial examples for the eight speed signs that we consider in this work (cf. Figure~\ref{fig:scenario_1}).

\begin{table*}[t]
\setlength{\arraycolsep}{2.5pt}
	\footnotesize
  \caption{Results for the five attack scenarios. Attack success rate and the average queries $Q$ under varying brightness conditions for fixed $k=192$ \textsf{MPs} and for varying number $k$ of \textsf{MPs} for a fixed brightness of 2000 lux. Here, $l=2$.}
	\centering 
	\scalebox{0.9}{$\begin{array}{cc|rrrr|rrrr|rrrr|rrrr||rrrr}
     \toprule
          &     & \multicolumn{16}{c||}{\textbf{Two-Stage Architecture}} & \multicolumn{4}{c}{\textbf{Single-Stage Architecture}} \\\cmidrule{3-22}
    & & \multicolumn{12}{c|}{\textbf{Targeted}} & \multicolumn{4}{c||}{\textbf{Untargeted}} & \multicolumn{4}{c}{\textbf{Hide}} \\\cmidrule{3-22}
    &&  \multicolumn{4}{c|}{\textbf{Scenario 1}} & \multicolumn{4}{c|}{\textbf{Scenario 2}} & \multicolumn{4}{c|}{\textbf{Scenario 3}} & \multicolumn{4}{c||}{\textbf{Scenario 4}} & \multicolumn{4}{c}{\textbf{Scenario 5}}    \\
  &  &  \multicolumn{4}{c|}{\text{Any speed $\rightarrow$ Lowest speed}} & \multicolumn{4}{c|}{\text{Any speed $\rightarrow$ Highest speed}} & \multicolumn{4}{c|}{\text{Stop $\rightarrow$ Any speed}} & \multicolumn{4}{c||}{\text{Any sign  $\rightarrow$ Any sign}} & \multicolumn{4}{c}{\text{Sign $\rightarrow$ No sign}}    \\\cmidrule{3-22}
   & &  \multicolumn{2}{c}{\textbf{GTSRB-CNN}}  &  \multicolumn{2}{c|}{\textbf{LISA-CNN}} & \multicolumn{2}{c}{\textbf{GTSRB-CNN}}  &  \multicolumn{2}{c|}{\textbf{LISA-CNN}} & \multicolumn{2}{c}{\textbf{GTSRB-CNN}}  &  \multicolumn{2}{c|}{\textbf{LISA-CNN}}&  \multicolumn{2}{c}{\textbf{GTSRB-CNN}}  &  \multicolumn{2}{c||}{\textbf{LISA-CNN}}  &  \multicolumn{2}{c}{\textbf{YOLOv8}}  &  \multicolumn{2}{c}{\textbf{Faster-RCNN}} \\  
  &              & \textsf{ASR} & Q & \textsf{ASR} & Q &\textsf{ASR} & Q &\textsf{ASR} &Q &  \textsf{ASR} & Q & \textsf{ASR} &Q & \textsf{ASR} &Q & \textsf{ASR} &Q & \textsf{ASR} &Q & \textsf{ASR} &Q\\\midrule
\multirow{6}{*}{\rotatebox[origin=c]{90}{\textbf{Lux}}}& 10 & 96.10 & 127.90 & 100.0 & 16.33 & 99.90 & 28.31 & 100.0 & 31.11 & 76.42 & 376.13 & 98.49 & 91.05 & 95.07 & 68.31 & 97.95 & 71.56 & 100.0 & 3.23 & 100.0 & 4.97\\
& 1000 & 95.62 & 135.36 & 100.0 & 15.02 & 99.79 & 28.92 & 100.0 & 31.33 & 73.42 & 400.05 & 98.42 & 103.76 & 94.88 & 70.62 & 97.81 & 77.68 & 100.0 & 3.21 & 100.0 & 4.38\\
& 2000 & 92.48 & 177.03 & 100.0 & 22.80 & 99.58 & 32.10 & 100.0 & 45.41 & 69.17 & 459.4 & 97.32 & 152.76 & 93.58 & 86.32 & 96.42 & 103.58 & 100.0 & 8.79 & 100.0 & 11.18\\
& 3000 & 81.43 & 300.15 & 100.0 & 39.13 & 95.94 & 89.87 & 99.51 & 70.58 & 57.17 & 589.18 & 90.66 & 297.81 & 89.02 & 146.33 & 93.86 & 167.98 & 99.16 & 19.33 & 98.53 & 39.07\\
& 4000 & 41.05 & 671.71 & 100.0 & 127.47 & 62.71 & 456.30 & 97.57 & 177.30 & 15.08 & 907.48 & 59.34 & 576.84 & 67.91 & 367.57 & 81.71 & 355.62 & 95.40 & 79.34 & 91.91 & 167.60\\
& 5000 & 4.67 & 964.87 & 62.80 & 585.45 & 10.42 & 922.86 & 56.80 & 644.25 & 1.25 & 995.47 & 25.76 & 876.95 & 28.56 & 747.15 & 40.75 & 725.77 & 84.52 & 301.02 & 56.25 & 596.34\\
\midrule
\multirow{6}{*}{\rotatebox[origin=c]{90}{\textbf{Patches ($k$)}}}& 16 & 57.14 & 559.79 & 99.39 & 70.66 & 87.19 & 260.26 & 99.03 & 109.45 & 59.08 & 490.19 & 87.71 & 287.74 & 86.88 & 172.60 & 87.93 & 234.34 & 98.33 & 55.33 & 96.69 & 79.44\\
& 32 & 76.76 & 363.14 & 100.0 & 35.70 & 95.10 & 135.89 & 100.0 & 58.03 & 63.50 & 441.24 & 95.67 & 175.91 & 90.60 & 121.12 & 94.59 & 150.15 & 99.16 & 22.15 & 100.0 & 23.36\\
& 64 & 88.67 & 218.16 & 100.0 & 24.06 & 98.44 & 71.78 & 100.0 & 39.42 & 67.00 & 418.08 & 97.53 & 127.67 & 92.84 & 96.76 & 96.56 & 105.23 & 99.16 & 13.83 & 100.0 & 6.37\\
& 96 & 90.95 & 192.13 & 100.0 & 20.01 & 98.65 & 55.78 & 100.0 & 32.84 & 68.25 & 427.06 & 97.87 & 120.78 & 93.67 & 87.65 & 96.63 & 97.19 & 100.0 & 9.81 & 100.0 & 5.83\\
& 128 & 92.57 & 172.00 & 100.0 & 17.71 & 98.85 & 47.11 & 100.0 & 37.61 & 68.33 & 438.48 & 97.87 & 128.01 & 93.67 & 86.66 & 97.15 & 91.65 & 100.0 & 9.15 & 100.0 & 5.21\\
& 192 & 92.48 & 177.03 & 100.0 & 22.80 & 99.58 & 32.10 & 100.0 & 45.41 & 69.17 & 459.40 & 97.32 & 152.76 & 93.58 & 86.32 & 96.42 & 103.58 & 100.0 & 8.79 & 100.0 & 11.18\\
   \bottomrule
	\end{array}$}
	\label{tab:targeted_asrs_light}
\end{table*}

\vspace{0.25 em}
To avoid bias toward specific geographic regions, we included traffic signs from both Europe (via the GTSRB dataset) and North America (via the LISA dataset). This approach ensured our analysis captured a diverse range of signs---with varying shapes, colors, and sizes---across both digital and physical experiments. More precisely, for GTSRB, we relied on the stop sign and speed limits of 20, 30, 50, 60, 70, 80, 100, and 120 km/h, with 150 samples for each class. Analogously for LISA, we opted to take all samples that are available for the selected classes due to the generally smaller dataset---here, we used speed limits 30, 35, and 45 to map to speed limit 25, speed limits 25, 30, and 35 to map to the highest available speed limit of 45, and mapped the stop sign to speed limits 25, 30, 35, and 45, for the three scenarios, respectively. 

Our results (cf. \Cref{tab:targeted_asrs_light}) indicate that our proposal consistently obtains a high \asr across all three targeted attack scenarios, datasets, number of \manypixel $k$, and strength of the ambient lighting, with up to $96.1\%$ and $100\%$ in Scenario 1, $99.9\%$ and $100\%$ in Scenario 2, and up to $76.42\%$ and $98.49\%$ in Scenario 3, for GTSRB and LISA, respectively. 
For LISA, we observe a slower decline in \asr, which we attribute to its fewer classes and their size in contrast to GTSRB. 

Our results also suggest that the targeted class flip from a stop sign to any speed sign, i.e., Scenario 3, is the most challenging to achieve, likely due to the dissimilarity between the two sign types.  In contrast, the scenarios involving similar signs, i.e., Scenarios 1 and 2, appear to be easier to realize in terms of higher \asr and fewer required queries. For GTSRB and Scenario 2, we obtain a high \asr $>99\%$ and a low query count of $28.3$ queries on average. For Scenario 1, we also observe a high \asr of up to $96.1\%$ at $127.9$ queries.

\vspace{0.25 em} \noindent \textbf{Impact of ambient light intensity: }
We now evaluate the performance of our proposal on the GTSRB and LISA datasets in the presence of a light source of varying intensity, ranging from 10 to 5000 lux, simulating a range from a dark to a brightly lit outdoor environment. Our results for this ablation are included in the upper half of \Cref{tab:targeted_asrs_light}. Here, we measure the \asr and the average consumed queries required in our approach for a combination of lux $\in \{10, 1000, 2000, 3000, 4000, 5000\}$ and $k=192$ (for the reasoning why, see next paragraph).

In the case of GTSRB, we mostly observe \asr of more than $90\%$ for lux values below 2000, except for the most challenging Scenario 3, where we reach an \asr around $70\%$. In contrast, LISA exhibits high success rates of more than $\sim90\%$ for values lower than $3000$ lux. For all three scenarios, we find the best lux setting for both datasets at $10$ lux for an \asr of $96.1\%/99.9\%/76.42\%$ at an average of $127.9/28.31/376.13$ consumed queries for GTSRB. In the case of LISA, we observe $100\%/100\%/98.49\%$ at $16.33/31.11/91.05$ average consumed queries for LISA. 
\begin{figure}[tb]
     \centering
      \includegraphics[width=0.75\linewidth]{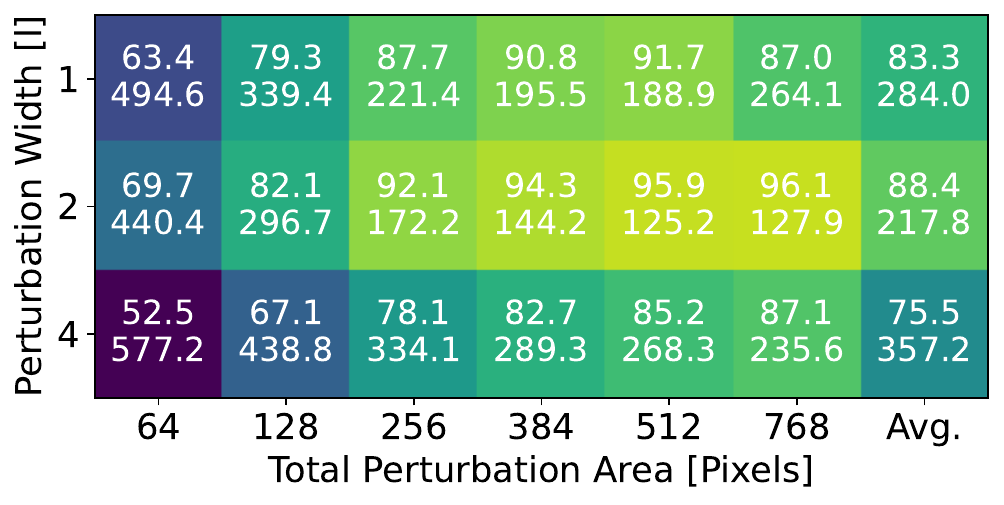}
  \caption{Perturbation width $l$ vs. amount of perturbed pixels on GTSRB for Scenario 1 with the average in the last column. Each cell contains the \asr and query count.}
\label{fig:results_width_ablation}
\end{figure}
The higher complexity for optimizing on GTSRB is also evidenced by the generally lower \asr and higher number of consumed queries compared to LISA, which confirms our previous observations. 
A core strength of our proposal lies in the modest number of required queries for convergence toward a successful \emph{targeted} adversarial example, e.g., $127.9$ queries for $10$ lux in Scenario 1 in GTSRB. 

\vspace{0.25 em} \noindent \textbf{Impact of number of \manypixel $\mathbf{k}$: }
In the lower half of \Cref{tab:targeted_asrs_light}, we vary the number $k$ of \textsf{MP}s between $16-192$ (out of the maximum of $256$ \textsf{MP}s) on the three proposed scenarios in a bright environment of 2000 lux. Recall that $k$ impacts the area covered by the perturbation. Generally, we observe that a varying $k$ can significantly boost the \asr by $\sim35\%$ and reduce the consumed queries by $\sim70\%$. In contrast, the attack success in the second and third scenarios is boosted by up to $12\%$. In the case of GTSRB, we observe that a minimum of $k=96$ is required to obtain an $\textsf{ASR}$ of $>90\%$ for the first two scenarios, reaching its maximum at $k=192$ at approximately $93\%/100\%$ for Scenarios 1 and 2, while reaching $70\%$ in Scenario 3. Generally, we observe that $k=192$ strikes a strong tradeoff between \asr and the query budget in all scenarios. 

\begin{table}[tb]
\footnotesize
	\caption{Detailed results for Scenario 1 for GTSRB with 2000 lux and $k=192$ used to compute the average \asr in~\Cref{tab:targeted_asrs_light}.}
	\centering 
	$\begin{array}{c|l}
     \toprule
    \text{Sign mapping} &  \asr \\\midrule
   30\rightarrow20 & 99.30   \\ 
50\rightarrow20 & 91.33   \\
60\rightarrow20 & 81.33   \\
70\rightarrow20 & 99.33   \\ 
80\rightarrow20 & 92.00   \\
100\rightarrow20 & 88.67   \\ 
120\rightarrow20 & 95.33   \\\bottomrule
	\end{array}$
 \label{tab:detailed_scenarios}
\end{table}

\vspace{0.25 em} \noindent \textbf{Impact of size of \manypixel $l$: }
Recall that the size $l$ of a \manypixel is directly proportional to the maximum number of $k$ \manypixel we can perturb. 
Since the number of \manypixels{} is bounded by the constant size of our samples $w=h=32$, changing the size of an \manypixel results in an upper bound on the total number of \textsf{MP}. For instance, consider the following configurations that all have the same number of underlying pixels while exhibiting a different number of \manypixels:\vspace*{.25em}
\begin{itemize}
    \item $l=1,k=256 \rightarrow \frac{32}{1} \times \frac{32}{1} = 1024$ \manypixel positions. \\$l^2 \times k = 1^2 \times 256 = 256$ perturbed pixels. 
    \item $l=2,k=64 \rightarrow \frac{32}{2} \times \frac{32}{2} = 256$ \manypixel positions. \\$l^2 \times k = 2^2 \times 64 = 256$ perturbed pixels. 
    \item $l=4,k=16 \rightarrow \frac{32}{4} \times \frac{32}{4} = 64$ \manypixel positions. \\$l^2 \times k = 4^2 \times 16 = 256$ perturbed pixels. 
\end{itemize}%
Therefore, we opted not to evaluate against $k$ but instead to benchmark against the number of perturbed pixels (see the above example). Our results are shown in \Cref{fig:results_width_ablation} for a targeted attack on GTSRB for Scenario 1. We observe an average \asr of $83.3\%$ at $l=1$, which increases to $88.4\%$ for $l=2$ and $75.5\%$ at $l=4$. For the standard case of $l=1$, i.e., in the case of pixel-wise perturbations, we observe the highest performance at $k=512$ with an \asr of $91.7\%$, which again decreases for a larger number of perturbed pixels, i.e., $768$, down to $87.0\%$. In contrast, we observe that a larger size $l$ is favorable due to the better performance of the attack. Particularly, we see that $l=2$ strikes the best tradeoff between the size of a \manypixel and the resulting available number of \manypixel $k$, where we observe the highest average \asr of $88.4\%$ and the highest overall \asr of $96.1\%$ for a total of $768$ perturbed pixels. Analogously to the \asr, we also observe a minimum of $217.8$ consumed queries on average for this configuration. This number of perturbed pixels results in a value of $k=192$ for $l=2$ (see previous paragraph).

\vspace{0.25 em} \noindent \textbf{Impact of sign choice: }Our approach is inherently general and does not exploit specific traffic sign shapes, colors, or textures. Notably, the choice of sign pairs has only a minor effect on \asr, as detailed in~\Cref{tab:detailed_scenarios}.

\begin{table*}[t]
	\footnotesize
  \caption{Results for a targeted attack on a yield and priority road traffic sign. \asr and average queries $Q$ under varying brightness conditions for fixed $k=192$ \textsf{MPs} and for varying number $k$ of \textsf{MPs} for a fixed brightness of 2000 lux. Results for real-world experiments are shown in ~\Cref{sec:real_world_exp}.
  }
	\centering 
\setlength{\arraycolsep}{2.5pt}
    \scalebox{0.83}{$\begin{array}{c|cccccc|cccccc|cccccc|cccccc}
     \toprule
& \multicolumn{12}{c|}{\includegraphics[height=.04\linewidth]{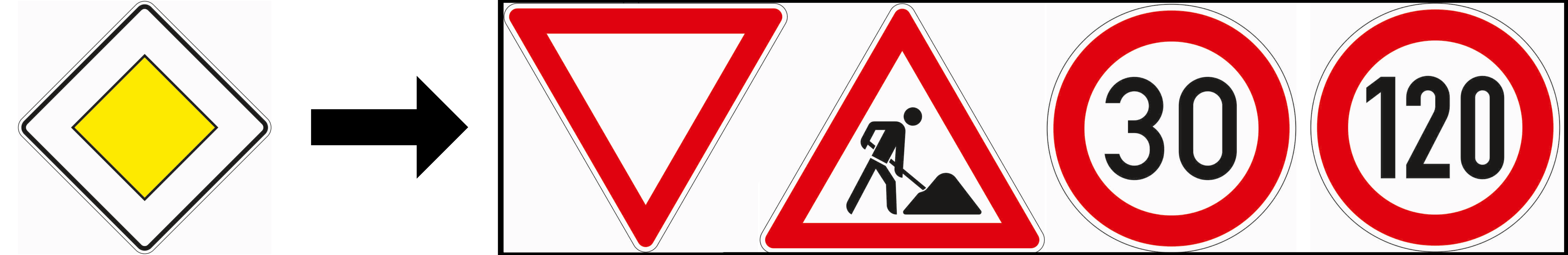}} & \multicolumn{12}{c}{\includegraphics[height=.04\linewidth]{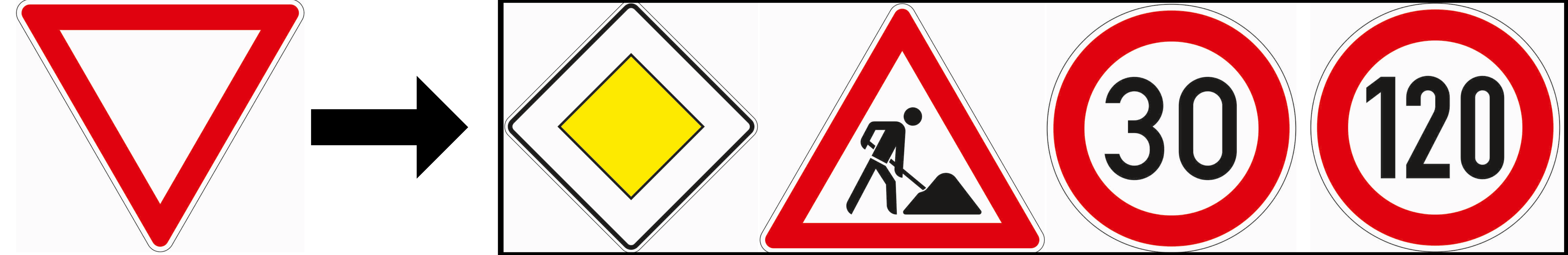}}\\\cmidrule{2-25}
&\multicolumn{6}{c|}{\textbf{Lux}} &\multicolumn{6}{c|}{\textbf{Patches ($k$)}} &\multicolumn{6}{c|}{\textbf{Lux}} &\multicolumn{6}{c}{\textbf{Patches ($k$)}} \\
&10 & 1000 & 2000 & 3000 & 4000 & 5000 & 16 & 32& 64& 96& 128& 192 &10 & 1000 & 2000 & 3000 & 4000 & 5000 & 16 & 32& 64& 96& 128& 192\\\midrule

\asr & 96.0 & 98.0 & 95.0 & 79.0 & 38.0 & 20.0 & 66.0 & 89.0 & 94.0 & 95.0 & 94.0 & 95.0 &83.0 & 85.0 & 81.0 & 70.0 & 46.0 & 12.0 & 39.0 & 58.0 & 81.0 & 76.0 & 77.0 & 81.0  \\
Q  & 137.2 & 155.9 & 193.9 & 358.4 & 693.5 & 822.2 & 476.8 & 231.4 & 161.3 & 141.3 & 155.6& 193.9 & 330.8 & 313.1 & 365.1 & 493.6 & 690.6 & 918.2 & 740.1 & 588.0 & 380.7 & 404.7 & 388.0 & 365.1 \\
    \bottomrule
	\end{array}$}
    \label{tab:dissimilar_signs}
\end{table*}

To confirm this intuition, we further extended our experiments beyond Scenarios 1–3---which already feature a diverse range of signs---to include several more dissimilar pairs. Specifically, we performed a targeted attack on the GTSRB priority road sign, aiming to misclassify it as a yield sign, a road construction sign, and a speed limit sign (30/120 km/h). As shown in~\Cref{tab:dissimilar_signs}, our method achieved a strong \asr of up to 98\% and an average of around 300 consumed queries. In addition, we consider a yield sign as a source and aim to classify it as a priority road, a road construction, and a speed limit sign (30/120 km/h) and obtain an \asr of up to 85\% with an average of 500 consumed queries.

\subsection{Untargeted Attacks in the Digital Domain}
\label{untargeted_eval}
We now move our focus towards an untargeted scenario.

\vspace{0.25 em}\noindent \textbf{Scenario 4 (Service disruption).} We use ``any'' sign as the source class and ``any'' sign as the target class. A class flip here can lead to a sudden stop, acceleration, or any other behavior triggered by a specific sign.

Our results for Scenario 4 are shown in the fourth column of \Cref{tab:targeted_asrs_light} for various brightness levels and a varying number of \manypixel. When compared to its targeted counterpart (cf. \Cref{targeted_eval}), we observe a less steep decline in \asr for the untargeted setting, combined with a slower increase in the number of queries for GTSRB, while both the ASR and number of queries for LISA are relatively similar. This trend highlights the increased difficulty in mounting targeted attacks compared to their untargeted counterparts.

In the case of GTSRB, we observe that a minimum of $k=32$ is required to obtain an $\textsf{ASR}$ of $\sim90\%$. Subsequent increases to $k=64$ result in a further boost of $\textsf{ASR}$ by $3\%$, which only marginally increases beyond that for larger values of $k$. We consistently observe \asrs of more than $\sim90\%$ for lux values below 3000, reaching $95.07\%$ at the lowest ambient lighting of 10 lux at just $68.31$ consumed queries for GTSRB. In contrast, we observe an \asr of $97.95\%$ and $71.56$ consumed queries for LISA. 
With a comparable \asr, we note that the number of queries required in LISA is slightly higher than that required in GTSRB. We contrast this to the previous trend, for which LISA performed better in terms of \asr and consumed queries, and attribute this to the fact that our sample size for this untargeted scenario is larger than the targeted scenarios before. 

\begin{table}[t]
\footnotesize
	\caption{\asr for various surrogate and target architectures of varying complexity. The bold diagonal elements indicate the \asr when surrogate and target architectures are identical.}
\centering 
\setlength{\arraycolsep}{1.25pt}
$\begin{array}{c|lcccc|lcc}
   \toprule
                                         &              & \multicolumn{4}{c|}{\textbf{Two-Stage}}  & & \multicolumn{2}{c}{\textbf{Single-Stage}} \\\midrule
     &    \multirow{2}{*}{{{\textbf{Target} $\rightarrow$}}}             & \multirow{2}{*}{{CNN}}  &  \multirow{2}{*}{\shortstack{Res-\\Net50}} & \multirow{2}{*}{{\shortstack{Swin-\\Trans.}}} &   \multirow{2}{*}{\shortstack{Conv-\\NeXt}} & \multirow{2}{*}{{{\textbf{Target} $\rightarrow$}}} & \multirow{2}{*}{{YOLOv8}} & \multirow{2}{*}{\shortstack{Faster-\\RCNN}} \\
   &  & & & &\\\midrule
    \multirow{4}{*}{\rotatebox[origin=c]{90}{\textbf{Surrogate}}} &                                                     \text{CNN} & \mathbf{95.16} & 74.88 & 57.77 & 57.58 & \multirow{2}{*}{{YOLOv8}} & \multirow{2}{*}{$\mathbf{100.00}$} & \multirow{2}{*}{$93.72$}\\ 
     &                                                   \text{ResNet50}    & 71.07 & \mathbf{97.58} & 55.81 & 56.37 \\ 
     &                                                   \text{SwinTrans.}   & 72.56 & 75.63 & \mathbf{97.12} & 66.98 & \multirow{2}{*}{\shortstack{Faster-\\RCNN}} & \multirow{2}{*}{$93.38$} & \multirow{2}{*}{$\mathbf{100.00}$}\\
     &                                                   \text{ConvNeXt}     & 72.84 & 77.40 & 68.84 & \mathbf{97.21} \\\bottomrule
\end{array}$
	\label{tab:transfer}
\end{table}

\vspace{0.25 em} \noindent \textbf{Blackbox transferability:}\label{sec:transfer}
To confirm that our approach is also effective on other architectures and to model an adversary without oracle access to the model, we now assess the transferability of our scheme to different architectures on the GTSRB dataset. 
Here, we use models of increasing complexity as surrogate models and generate the adversarial perturbations, which we subsequently evaluate on the target architecture for Scenario 4. We consider the following architectures (with the respective number of weights): GTSRB-CNN ($\sim16.5$M), ResNet-50~\cite{heDeepResidualLearning2016} ($\sim25.5$M), SwinTransformer~\cite{liu2021swin} ($\sim87.7$M), and ConvNeXt~\cite{liu2022convnet} ($88.5$M). Our experiments are conducted for 10 lux, $k=192,l=2$, and a query budget of $Q=1000$. Unlike our previous experiments, where we stopped the attack once an adversarial example was found, we utilized the entire query budget here to more accurately assess the robustness of the perturbation.
Our results, summarized in \Cref{tab:transfer}, show that the success of our attack is independent of the underlying model architecture. Specifically, for the same surrogate and target models, we consistently achieve success rates of over $95\%$. We observe higher transferability rates from the more complex architectures towards the simpler ones, i.e., the first column shows an average transferability of $\sim 72\%$ towards the simplest architecture GTSRB-CNN. On the other hand, when using GTSRB-CNN as the surrogate model, we observe a transferability of $\sim 63\%$ to the more complex architectures.

\vspace{0.25 em} \noindent \textbf{Comparison with Related Work:}\label{sec:related_work_comparison}
We now compare our approach against the state-of-the-art methods of~\cite{weiPhysicallyAdversarialInfrared2023, weiHOTCOLDBlockFooling2023, lovisottoSLAPImprovingPhysical} using the GTSRB dataset.
The former two attacks are black-box methods based on transferability and gradient-free particle swarm optimization \cite{488968}, respectively, while the latter is a white-box method with direct model access and, as such, requires the availability of model gradients. 

We adapt~\cite{weiPhysicallyAdversarialInfrared2023, weiHOTCOLDBlockFooling2023} to a two-stage pipeline and optimize our loss functions to evaluate the effectiveness of their shape-generation strategies in Scenario 4.
We instrument our approach with $k=192$ and $l=2$, as determined in the aforementioned ablation study, and apply our infrared transformation with a brightness of $10$ lux. 

Our results are depicted in \Cref{tab:asrs_rw}\footnote{We could unfortunately not compare with \cite{weiUnifiedAdversarialPatch2023, satoInvisibleReflectionsLeveraging2024} due to the unavailability of their source code.}. We find that our proposal results in a remarkably higher \asr by at least {12.5\%} and a lower amount of queries, by up to 65\%, compared to~\cite{weiPhysicallyAdversarialInfrared2023,weiHOTCOLDBlockFooling2023}, even though~\cite{weiPhysicallyAdversarialInfrared2023} is a white-box method with direct access to model gradients. 

To compare against~\cite{lovisottoSLAPImprovingPhysical}---a projector-based attack in the \emph{visible} light spectrum, we generate perturbations at $120$ lux for 100 stop signs\footnote{Note that a comparison with the full GTSRB was not feasible as the projection model of ~\cite{lovisottoSLAPImprovingPhysical} is sign-specific and is only available for a stop sign.} sampled from GTSRB and obtain an \asr of $100\%$ at an average of just $2.42$ queries. Our results are on par with the results in \cite{lovisottoSLAPImprovingPhysical}; however, our scheme does not require the generation of individual projection models and saves considerable effort in generating adversarial examples. 

\begin{table}[t]
	\footnotesize
 	\caption{Comparison against state-of-the-art w.r.t. $\mathsf{ASR}$, average queries $Q$ on GTSRB, and time it takes for deployment.}
	\centering 
	$\begin{array}{c|cc|c}
     \toprule
	&    \text{Shapes \& Location\cite{weiPhysicallyAdversarialInfrared2023}}  & \text{HotNCold \cite{weiHOTCOLDBlockFooling2023}} & \textbf{Ours} \\
        \midrule
 \asr   & 74.6  & 82.6 & \mathbf{95.07}\\ 
 Q  &  200.0  & 200.0 &\mathbf{68.31} \\ 
  \text{Time}   & \sim 5\text{ min}  & 30 \text{ min} & \sim \mathbf{50}\textbf{s}\\ 
   \bottomrule
	\end{array}$
	\label{tab:asrs_rw}
\end{table}

\subsection{Perturbation Attacks in the Physical World} \label{sec:real_world_exp}

\begin{figure}[t]
     \centering
            \subfloat[Indoor setting.\label{fig:indoor_environment}]{\centering\includegraphics[width=.43\linewidth]{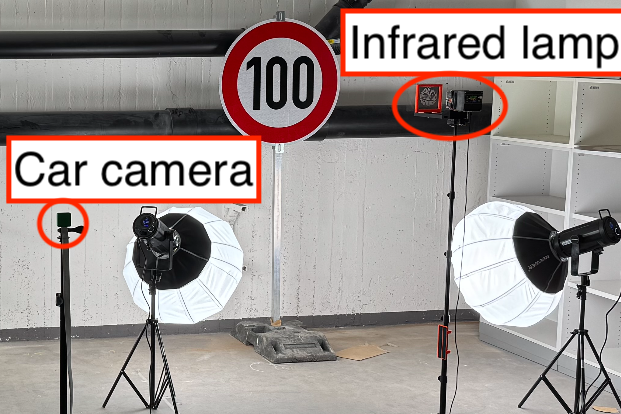}}
            \hspace{0.1em}
            \subfloat[Outdoor setting.\label{fig:outdoor_environment}]{\centering\includegraphics[width=0.43\linewidth]{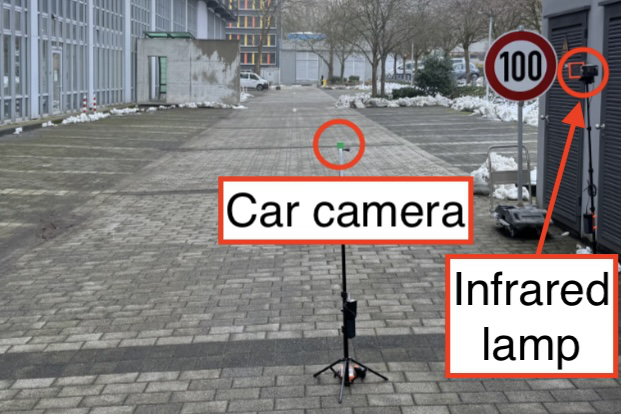}}
            \caption{Experimental environments with 1000 lux (avg.) on the sign surface.}
            \label{fig:exp_environments}
\end{figure}

We now proceed to evaluate our approach in the real world. In our experiments, we directly perturb the $w=h=224$ large images using a square \manypixel with $l=14$ and apply the aforementioned EOT transformations (cf. \Cref{sec:real_world_pert}) while enforcing a query budget of $Q=2500$ to make the perturbations more robust. 
In practice, generating a single perturbation takes approximately four minutes and needs to be performed only once before deployment. Because these attacks transfer effectively across classifiers (cf.~\Cref{untargeted_eval}), the adversary does not need to interact directly with the target classifier in the vehicle. Due to the larger image size, we opted to rely on this (large) \manypixel size to facilitate the recognition by a camera. Here, we select one representative class mapping for each introduced scenario and devise ten dedicated perturbations, i.e., we average the success of each scenario over ten different perturbations. An example of Scenario 1 is the targeted class flip from speed limit 100 to speed limit 30.  

\vspace{0.25 em} \noindent \textbf{Setup \& Hardware: }
We performed our experiments using two different cameras with CMOS sensors, which are commonly found in product families used in autonomous driving or commercial traffic sign recognition systems, such as Baidu Apollo. For most experiments, we use (1) Raspberry Pi Camera Module 3 without infrared filters based on a Sony IMX708 sensor with a focal length of $4.74$mm (similar to Leopard Imaging \texttt{LI-USB30-IMX728-GMSL3-070H}). In another dedicated test, we also relied on (2) Leopard Imaging \texttt{LI-USB30-AR023ZWDR} (using an OnSemi \texttt{AR023ZWDR} sensor) with a focal length of $6$mm, which has also been used in other works~\cite{satoInvisibleReflectionsLeveraging2024}. Both cameras have been connected to a Raspberry Pi Model 4. 
To broaden the range of ambient light intensity conditions, we used a powerful 12W 808nm infrared light source in all our experiments\footnote{In our initial tests, we used a 5W 850nm infrared lamp to successfully mount attacks up to an ambient light intensity of 300 lux.}. This allows us to produce clearly visible perturbations even under high ambient light levels of up to 1100 lux. Ambient light intensity is measured directly on the surface of the sign using a lux meter. 

We printed the previously generated perturbations on transparent off-the-shelf overhead projector film made from PET, costing around US\$0.1 per perturbation. This process is sufficiently precise for our purpose, as initial experiments on different versions of the same perturbation did not show any impact on the attack success rate.

\vspace{0.25 em} \noindent \textbf{Placement: }
For placing the camera and infrared light source in our experiments, we assume a real-world setting mimicking a traffic setting (cf. \Cref{fig:system_model}), in which we place our sign on the right side of the road. We place the infrared light source at a fixed position opposite the sign at a distance of 2 meters, considering one-shot attackers. This choice is reasonable, as traffic signs are typically located on the side of the street, which is also the only practical place to position a light source (e.g., on bridges, alternative side placements may not be feasible).
The projection was manually aligned once using the live camera feed prior to any experiment and remained unchanged throughout experiments. 
The default position of our camera is located in the middle of the right driving lane at a distance of $4$ meters (longitudinal) and $2$ meters (lateral) to the left of the sign, at a default viewing angle of $\sim25\degree$.

\begin{figure}[t]
     \centering
\resizebox{\linewidth}{!}{%
\includegraphics[width=\textwidth]{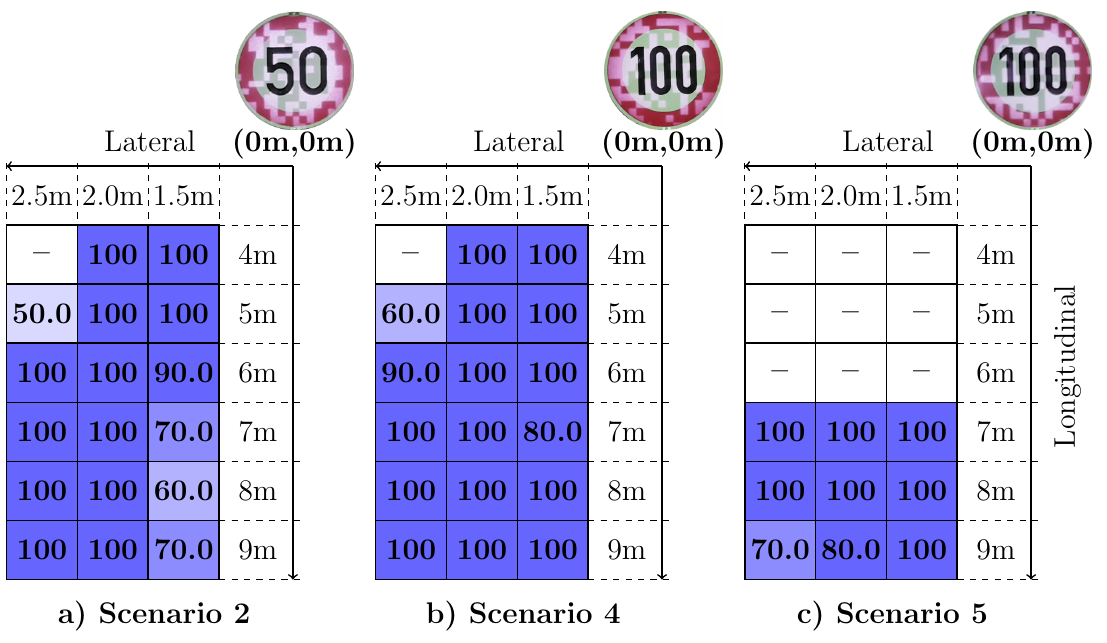}
}
\caption{Attack success rate for various scenarios at different camera positions in an indoor setting at a brightness of 1000 lux. We omitted the datapoint at (4m, 2.5m) because the sign was not fully within the camera's field of view.}
\label{tab:asrs_rw2}
\end{figure}

We first evaluate the success of our approach in a controlled and artificially lit indoor environment, i.e., a basement with bright natural video lighting (\Cref{fig:indoor_environment}), and then move into a more diverse outdoor scenario, i.e., a parking lot (\Cref{fig:outdoor_environment}). In both settings, we measure an average ambient lighting of 1000 lux. At all times, we verified the correct classification even in the presence of an infrared light spot (without a perturbation).

\vspace{0.25 em} \noindent \textbf{General Success:} We evaluate the performance of our approach in the previously introduced scenarios in both indoor and outdoor environments (cf. \Cref{tab:asrs_general_success}). To this end, we place the camera at the previously described distance and determine the \asr over ten different perturbations. In the indoor setting, we obtain an \asr of $100\%$ for Scenarios 1, 2, 3, and 4. 

In the outdoor environment, we observe success rates of $90\%$ and $80\%$ for the first two scenarios, respectively, while the last two scenarios maintain a success rate of $100\%$. In a separate experiment (conducted outdoors) on a yield sign flipping to a priority road sign, we observe an \asr of $70\%$. 

\vspace{0.25 em} \noindent \textbf{Different Angles, Distances, Cameras: } To assess the impact of real-world environments, such as spatial transformations introduced by angle and distance on the robustness of the perturbed signs and the success of our EOT transformations (cf. \Cref{sec:real_world_pert}), we conduct the following experiments: we place the camera at the default distance of $4$ meters of longitudinal and $2$ meters of lateral distance away from the sign and verify the success for various camera positions. Namely, we simulate different lane positions of the vehicle on the road by moving the camera laterally to the left and right by $0.5$ meters. We combine this with longitudinal distances between 4 and 9 meters in one-meter increments to verify the robustness of our approach, which results in diverse viewing angles between {$27\degree$ and $10\degree$}. These positions faithfully capture various real-world positions across different driving lanes and are also limited by the visibility of the sign on the camera. As shown in \Cref{tab:asrs_rw2}, we observe consistently high \asr, averaging over $90\%$ in almost all configurations. However, we observe a reduction in \asr for (5m, 2.5m) due to the slightly steeper viewing angle and increased distance (resulting in a less visible reflection). Another decrease in \asr is measured when increasing the longitudinal distance across a lateral distance of 1.5m due to the reflections of the infrared light source becoming more prominent on the sign (especially for the targeted Scenario 2).

\begin{figure}[tb]
     \centering
    \subfloat[Raspberry Pi Camera 3]{
    \centering\includegraphics[width=.23\linewidth, height=.23\linewidth]{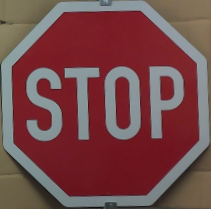}\includegraphics[width=.23\linewidth, height=.23\linewidth]{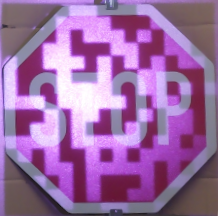}
    }
    \subfloat[Leopard Imaging \texttt{AR023ZWDR}]{
    \centering\includegraphics[width=.23\linewidth, height=.23\linewidth]{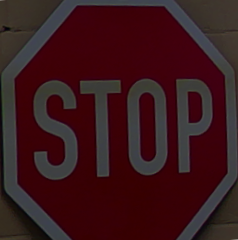}\includegraphics[width=.23\linewidth, height=.23\linewidth]{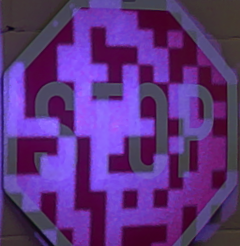}
    }
  \caption{Infrared perturbation captured with two different camera sensors.}
\label{fig:varying_cameras}
\end{figure}
\begin{figure}[t]
\centering\includegraphics[width=\linewidth]{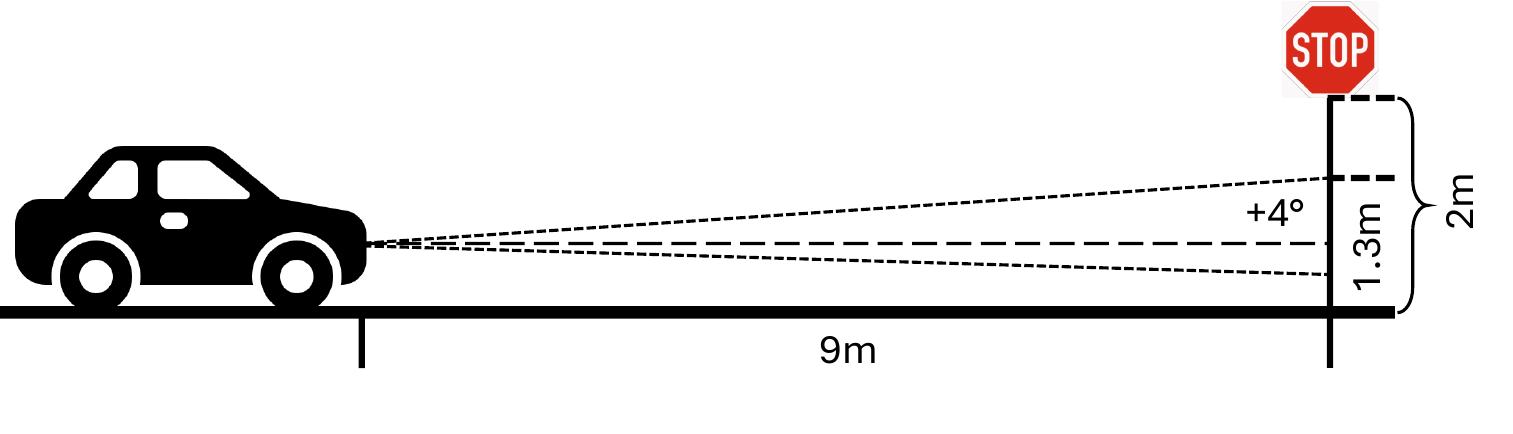}
  \caption{Impact of headlights (cf. ECE-R112~\cite{http://data.europa.eu/eli/reg/2014/112/oj}) at a distance of 9m on a traffic sign mounted at a height of 2m. The maximum permitted brightness on the surface of the sign is $\sim 22$ lux. Figure is to scale.}
\label{fig:headlight_schematic}
\end{figure}

To evaluate the transferability of our approach across different camera sensors, we also instrumented an additional camera, namely Leopard Imaging \texttt{AR023ZWDR} (cf. \Cref{fig:varying_cameras}), in the indoor setting. In this setting, we tested ten different perturbations for Scenario 1 and observed a high transferability rate of $90\%$. As the spectral sensitivity curves of CMOS camera sensors in the near-infrared part of the spectrum (800-1000nm) are highly similar, we expect our approach to also be effective against other sensors.

\vspace{0.25 em} \noindent {\textbf{Driving vehicle: } We now conduct moving vehicle experiments by driving past the perturbed sign in an outdoor environment. In our setup, the camera is mounted on the rear-view mirror, which corresponds to the typical height for front-facing camera systems in modern vehicles. Starting from a distance of 30 meters, we approach the sign at two different speeds: 10 km/h and 30 km/h. For safety reasons, we were unable to experiment at higher speeds. For each perturbation and across all five scenarios, we record a video and compute the \asr over all cropped frames (cf. \Cref{tab:asrs_general_success}). Our results demonstrate the practical effectiveness of our approach, achieving an \asr of up to $98\%$ and as low as $79.8\%$ at speeds of up to 30 km/h. Some fluctuations are observed, which we attribute to slight variations in the driving path and the fact that the experiments were conducted over several hours. 

\vspace{0.25 em} \noindent \textbf{Impact of headlights: } Headlights, particularly at dusk or night, create high-brightness conditions. 
To evaluate the impact of headlights on the robustness of our approach, we conducted outdoor tests under headlight illumination. We found no significant impact on our results within the tested range of 4–9m. %
As shown in ~\Cref{fig:headlight_schematic}, this resilience is mainly due to regulatory constraints designed to minimize glare for other drivers. Specifically, ECE-R112~\cite{http://data.europa.eu/eli/reg/2014/112/oj} [Figure B and Section 6.2.4] stipulates that headlight illumination at a height of 2 meters—where traffic signs are typically placed—must not exceed $\sim22$ lux at a distance of 9 meters.

\begin{table}[t]
	\footnotesize
\caption{\asr in the physical world in the indoor and outdoor environment for a given scenario.}
\centering 
	$\begin{array}{c|cccc|c}
     \toprule
                & \multicolumn{5}{c}{\textbf{Scenario}} \\\midrule
     \text{Environment} & \#1 & \#2 & \#3  & \#4 & \#5 \\\midrule
     \text{Indoor} & 100.0  &100.0 & 100.0  &100.0 & 100.0 \\
     \text{Outdoor} & 90.0  &80.0 & 100.0  &100.0 &100.0 \\\midrule
     \text{Moving (10 km/h)} & 99.4 & 93.7 & 96.3 & 84.8 & 79.8 \\
     \text{Moving (30 km/h)} & 98.0 & 90.0 & 84.5 & 84.4 & 85.7 \\
   \bottomrule
	\end{array}$
	\label{tab:asrs_general_success}
\end{table}

\section{Experiments on Single-Stage Architectures}\label{single-stage-eval}
We now shift our focus to single-stage architectures, which are typically used in object detection. 
Here, we conducted our experiments on the established Mapillary\cite{10.1007/978-3-030-58592-1_5} and GTSDB\cite{stallkampManVsComputer2012} datasets. Mapillary consists of 401 classes with traffic signs from all continents, while GTSDB contains 43 different German road sign classes, similar to the previously used GTSRB.
We train a YOLOv8 model\cite{glennjocherandayushchaurasiaandjingqiuUltralyticsYOLOv82023} on the Mapillary dataset with a reduced number of classes, i.e., European speed limit signs and stop signs, to achieve better performance (cf. \Cref{sec:back}), and obtain an mAP-50 of $64.9\%$. Additionally, we train a Faster-RCNN~\cite{renFasterRCNNRealTime2017} model on GTSDB and obtain an mAP-50 of $90.76\%$. 

\vspace{0.25 em} \noindent \textbf{Hiding Attacks (Scenario 5) in the Digital Domain:} In the setting of a single-stage architecture, the goal of the adversary is to ensure that the sign is no longer detected by the system. In other words, speed limits and other important signs, e.g., stop signs, are ignored by the traffic sign recognition system. 

We use 25 images per class for Mapillary, i.e., a total of 225 images, and the entire test set for the previously selected classes of GTSDB, while ensuring that we only select bounding boxes with more than $32\times32$ pixels.
As shown in \Cref{tab:targeted_asrs_light}, we obtain high \asrs at $k=192$ of $100\%$ for an average of $3.23/4.97$ queries for Mapillary and GTSDB, respectively. Even for the single-stage architectures, we measure high success rates and a lower amount of used queries at a higher value of $k$ and a lower ambient light level. 

To assess whether a perturbation generated for one architecture is also successful on another, we use the generated images for Mapillary on YOLOv8 and evaluate the success of a hiding attack on the Faster-RCNN model trained on GTSDB (and vice versa). As shown in \Cref{tab:transfer}, we measure a higher transferability of $\sim93\%$ compared to two-stage architectures.

\vspace{0.25 em} \noindent \textbf{Perturbation Attacks in the Physical World: } 
Analogously to the two-stage experiments, we place the camera at a distance of 7 meters and generate ten perturbations in Scenario 5. As shown in \Cref{tab:asrs_general_success}, we measure a success rate of $100\%$. Note that a larger initial distance is necessary for initial detection, as the dataset consists of more images with smaller signs at a distance rather than close-up signs.

In \Cref{tab:asrs_rw2}, we further vary the distance and angle between the camera and the traffic sign (starting from the initial distance of 7 meters). Our results consistently show an average success of $\sim95\%$. 

\section{Defenses against Infrared Perturbations} \label{sec:defenses}
Since infrared spectral filters impair camera performance in low-light conditions (cf. Appendix~\ref{infrared_films}), we now explore the solution space to defend against infrared perturbations and then present our defense, dubbed \emph{segmentation-based detection}.

\begin{table}[t]
\setlength{\arraycolsep}{3.5pt}
	\footnotesize
 	\caption{Impact of current defenses on the \textsf{CA} and \asr on GTSRB (digital domain) and on our experimental data (physical domain). $\uparrow$ (resp. $\downarrow$) indicates that values close to 100 (resp. 0) provide better results.} 
	\centering 
	$\begin{array}{c|cccc|c}
    \toprule
 & \multicolumn{4}{c|}{\textbf{Digital}} & \multicolumn{1}{c}{\textbf{Physical}}\\\midrule
 & \multirow{3}{*}{\shortstack{No\\defense}}  & \multirow{3}{*}{\shortstack{Spatial\\Smooth.\\(non-local)\cite{xuFeatureSqueezingDetecting2018}}} & \multirow{3}{*}{\shortstack{Spatial\\Smooth.\\(local)\cite{xuFeatureSqueezingDetecting2018}}}  & \multirow{3}{*}{\shortstack{Adv.\\Training\\\cite{goodfellowExplainingHarnessingAdversarial2015}}} & \multirow{3}{*}{\shortstack{\textbf{Ours}\\Segment.\\-based}}  \\
            & & & & &\\
            & & & & &\\\midrule
   \textsf{CA} \uparrow & 98.76 & 95.35 &{96.56}  & 98.67 & \mathbf{96.63}  \\ 
   \asr \downarrow & 95.16 & 67.72 &{61.77}& 62.89 & \mathbf{25.3} \\\bottomrule
	\end{array}$
	\label{tab:asrs_defended}
\end{table}

\subsection{Limitations of Current Defenses}
\vspace{0.25 em} \noindent \textbf{Spatial Smoothing \& Adversarial Training: }
First, we evaluate the impact of two popular defenses on our approach: the test-time spatial smoothing defense~\cite{xuFeatureSqueezingDetecting2018} and the popular (but costly) adversarial training~\cite{goodfellowExplainingHarnessingAdversarial2015}.

Local smoothing applies a median blur by replacing each pixel with the median of its neighbors, while non-local smoothing uses a larger region. Both aim to undo adversarial perturbations, following prior work~\cite{weiPhysicallyAdversarialInfrared2023, weiUnifiedAdversarialPatch2023}. Adversarial training strengthens test-time robustness by incorporating adversarial examples into the training process.

Our results for the strongest attacker, i.e., an infrared transformation for 10 lux, are included in \Cref{tab:asrs_defended}. We observe that all three defenses fail to fully mitigate our attack: test-time defenses~\cite{xuFeatureSqueezingDetecting2018} reduce \asr to $67.72\%$ (non-local) and $61.77\%$ (local), while adversarial training lowers it to only $62.89\%$.

\vspace{0.25 em} \noindent \textbf{Certified patch detection:} PatchCleanser~\cite{DBLP:conf/uss/0001MM22} is a certified defense that selectively masks portions of an image---if the mask covers an adversarial patch, the prediction of the classifier changes. In contrast, benign natural images are generally invariant to this mask. This does not apply to our use case of traffic-sign recognition as masking, e.g., a speed sign, creates an ambiguity of the underlying speed limit~\cite{satoInvisibleReflectionsLeveraging2024}. An additional requirement of PatchCleanser is that the mask must be larger than the used adversarial patch---we, however, perturb the entire sign with our perturbation. 

\vspace{0.25 em} \noindent \textbf{Infrared speckle detection:}
\cite{satoInvisibleReflectionsLeveraging2024} uses the characteristic speckle pattern of laser reflections for detection. While we also utilize infrared light, our approach uses an incoherent light source, i.e., \emph{not} a laser, and hence our perturbations do not exhibit a strong speckle pattern as required by~\cite{satoInvisibleReflectionsLeveraging2024}\textsuperscript{6}.

\vspace{0.25 em} \noindent \textbf{Spatio-temporal consistency:} 
When conducting evasion attacks in the real world, it has been shown that evading individual camera frames is not sufficient to successfully attack a system~\cite{shenSoKSemanticAI2024}. Indeed, by monitoring the spatio-temporal properties of objects, one can detect changes in bounding box size and classification over time~\cite{xuPhyScoutDetectingSensor2024, hanVisionGuardSecureRobust2024, 285459}\footnote{Notice that a comparison to~\cite{satoInvisibleReflectionsLeveraging2024, xuPhyScoutDetectingSensor2024,hanVisionGuardSecureRobust2024} is not possible since the source code has not been made available to us or cannot be extended to new attacks.}. These approaches typically rely on inconsistencies resulting from adversarial perturbations and can only be defeated when the model predictions are consistent ``enough'' over time, while also considering a model's natural error rate.
In our moving vehicle experiments (cf. \Cref{tab:asrs_general_success}), we obtain \asrs of up to $99.4\%$. Specifically, our targeted attacks are successful over most captured frames and therefore cannot be detected using such approaches. These results show a consistent targeted misclassification \emph{over the 163 frames of the video, with only one flickering frame scattered in between} (i.e., with an error rate of 0.6\%), which we attribute to the model's natural error rate due to motion blur. Importantly, as the majority of frames while approaching a sign are consistent, we believe that defenses based on spatio-temporal consistency will have a limited effect here. 

\begin{figure}[t]
     \centering
  \subfloat[Perturbed sign.\label{fig:perts_seg}]{\makebox[1.1\width][c]{\centering\includegraphics[width=.26\linewidth]{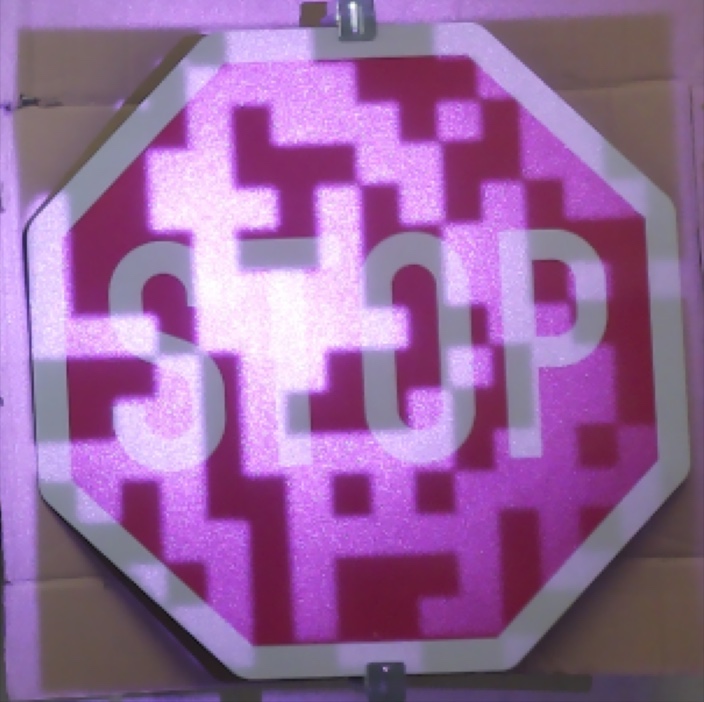}}}
  \hspace{0.1 em}
    \subfloat[Perturbed sign (segmented).\label{fig:seg_perts}]{\makebox[1.1\width][c]{\centering\includegraphics[width=.26\linewidth]{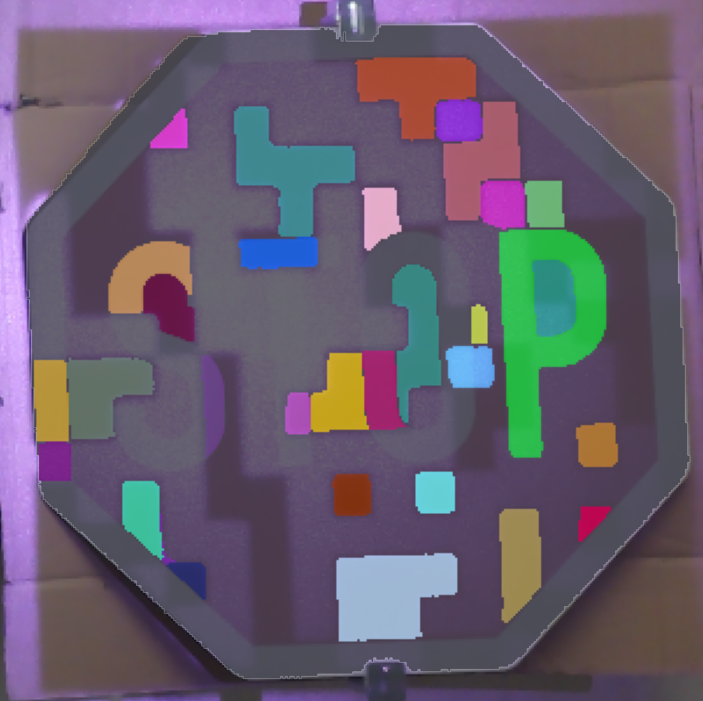}}}
     \hspace{0.1 em}
     \subfloat[Segmented sign.\label{fig:normal_seg}]{\makebox[1.1\width][c]{\centering\includegraphics[width=.26\linewidth]{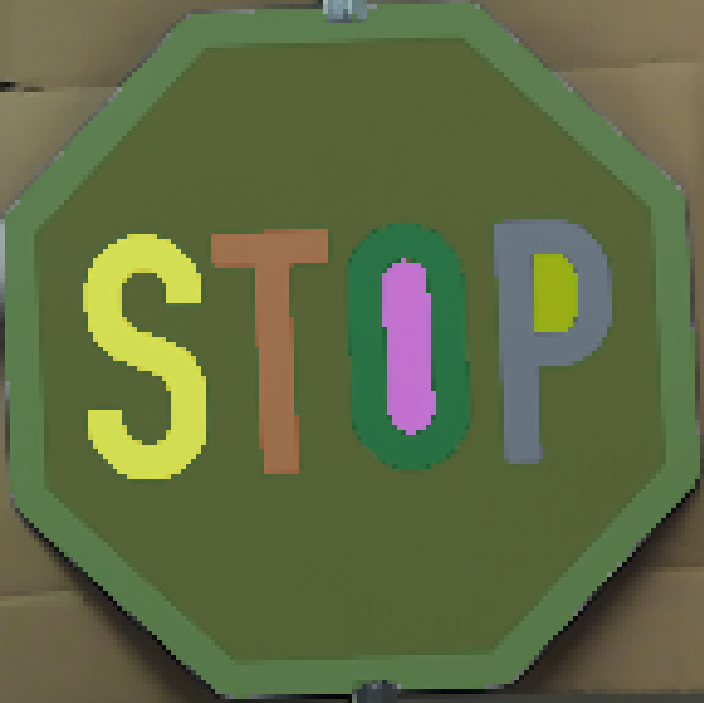}}}
  \caption{Output of our segmentation patched defense for a stop traffic sign.}
\label{fig:patch_segmentation}
\end{figure}

Some approaches like ~\cite{yuPhySenseDefendingPhysically2024} utilize object texture, behavior, and interactions with one another and focus specifically on detecting pedestrians and cars. This approach is not effective for traffic sign recognition as traffic signs have a similar texture, remain on fixed trajectories, and generally do not interact with other objects (like cars and pedestrians). 

\subsection{Our Proposal---Segmentation-based Detection}

We now propose a novel detection scheme specifically designed to thwart our attack. Our defense builds on the observation that our perturbations introduce a significant amount of additional shapes and edges into the image---considerably beyond the number of edges/shapes that are typically present in common traffic signs (cf. \Cref{fig:patch_segmentation}). 
More specifically, our defense measures the number of detected shapes in a given input image and compares it to an empirically derived threshold $\nu$, above which the image is considered adversarial.

To ensure robust and brightness-agnostic detection of shapes within the image, we utilize the Segment~Anything~{\cite{kirillovSegmentAnything2023} segmentation model with the ViT-L architecture to compute segmentation masks of an input image. This model $\mathcal{F}$ outputs a mask $m$ with $t$ pixels, i.e., $m=\{(x_0, y_0),\ldots,(x_t, y_t)\}$, for each of the $u$ detected shapes within an image, constituting the set $\mathcal{R}=\{m_0, \ldots, m_u\}$, i.e., $\mathcal{F}(x)=\mathcal{R}$. 

\vspace{0.25 em}\noindent \textbf{Dataset.} To evaluate our defense, we relied on a dataset consisting of 54 benign/unperturbed and 400 adversarial/perturbed images of traffic signs taken in static scenarios, as well as 476 and 1158 images taken from a moving vehicle, respectively (both obtained from our real-world experiments in ~\Cref{sec:real_world_exp}). 

\begin{table}[t]
\setlength{\arraycolsep}{3pt}
	\footnotesize
 	\caption{Evaluation of our segmentation-based detection scheme in the physical domain. $\uparrow$ (resp. $\downarrow$) indicates that values close to 100 (resp. 0) provide better results.}
	\centering 
	$\begin{array}{c|ccccc|c}
    \toprule
 & \multicolumn{5}{c|}{\textbf{Static}} & \multicolumn{1}{c}{\textbf{Moving}} \\\cmidrule{2-7}
 & \multirow{2}{*}{\shortstack{Indoor}}  & \multirow{2}{*}{Outdoor} & \multirow{2}{*}{\shortstack{Distance\\(Longitudinal)}}  & \multirow{2}{*}{\shortstack{Distance\\(Lateral)}} & \multirow{2}{*}{$\Sigma$} & \multirow{2}{*}{$\Sigma$} \\
 & & & & & \\\midrule
   \textsf{CA} \uparrow & 100 & 100 &100  & 100 & 100 & 96.63 \\ 
   \asr \downarrow & 2.49 & 2.56 & 1.63 & 1.65 & 2.05 & 25.3\\
  \bottomrule
	\end{array}$
	\label{tab:optimal_nu_all_data}
\end{table}

\vspace{0.25 em}\noindent \textbf{Determining $\mathbf{\nu}$: } We interpolated the threshold $\nu$ on the number of detected shapes, i.e., $|\mathcal{R}|$, experimentally based on the benign and perturbed traffic signs captured in our static experiments (diverse lighting conditions, distances), and from a moving vehicle.
The distribution of segmentation masks is presented in~\Cref{fig:histo}, supporting our initial hypothesis that benign images contain fewer detected shapes than adversarial ones, with only limited overlap between the two distributions. To evaluate detection performance across all possible threshold values $\nu$, we employ a receiver operating characteristic (ROC) curve, as shown in~\Cref{fig:roc}. Performance across different data partitions is summarized in~\Cref{tab:optimal_nu_all_data}.

\vspace{0.25 em}\noindent\textbf{Performance in static scenarios: }
For the optimal threshold $\nu=22$, we obtain a $\mainacc=100\%$ and an \asr as low as $2.05\%$ at an equal-error rate (EER) of $\sim2\%$ and an F1-score of $99\%$. 
Within the static scenarios, we observe no differences in performance between indoor and outdoor settings, nor any variations related to specific distances in latitude or longitude.

\vspace{0.25 em}\noindent\textbf{Performance under a moving vehicle: } To evaluate real-world performance in a dynamic setting, we utilize the aforementioned dataset---collected from a moving vehicle. While performance shows a slight decline relative to the static case, the impact on $\mainacc$ remains minimal, and our approach continues to outperform all other defenses by a substantial margin (see~\Cref{tab:asrs_defended}). Specifically, for a threshold of $\nu=9$, we achieve an $\asr$ of $25.3\%$ at an F1-score of $85\%$.
This slight performance degradation is likely due to the inclusion of images captured from the moving vehicle at distances of up to 30 meters, resulting in smaller object sizes that make segmentation more challenging.

\begin{figure}[tb]
     \centering
    \subfloat[Static setting.\label{fig:seg_static}]{\centering\includegraphics[width=0.47\linewidth]{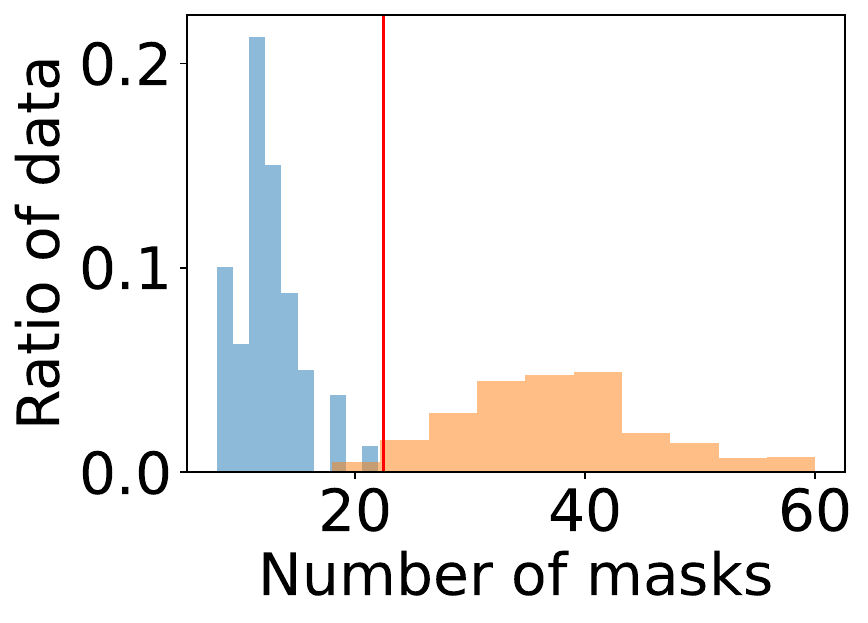}}
    \subfloat[Moving vehicle.\label{fig:seg_moving}]{\centering\includegraphics[width=0.47\linewidth]{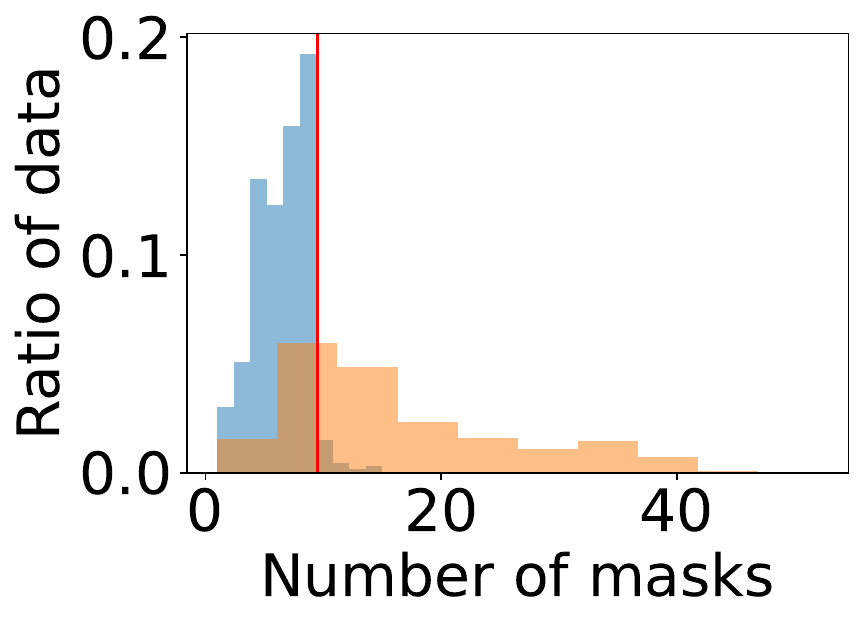}}
  \caption{Distribution of the segmentation masks for the benign (blue) and adversarial (orange) data for images taken in a static and a moving setting. The red line indicates the EER-optimal $\nu$.}
\label{fig:histo}
\end{figure}

\begin{figure}[t]
\subfloat[Static setting.\label{fig:roc_static}]{\centering\includegraphics[height=2.5cm]{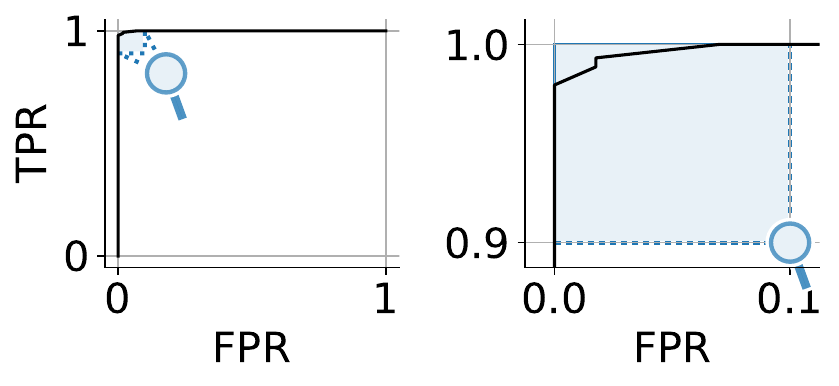}}
    \subfloat[Moving vehicle.\label{fig:roc_moving}]{\centering\includegraphics[height=2.35cm]{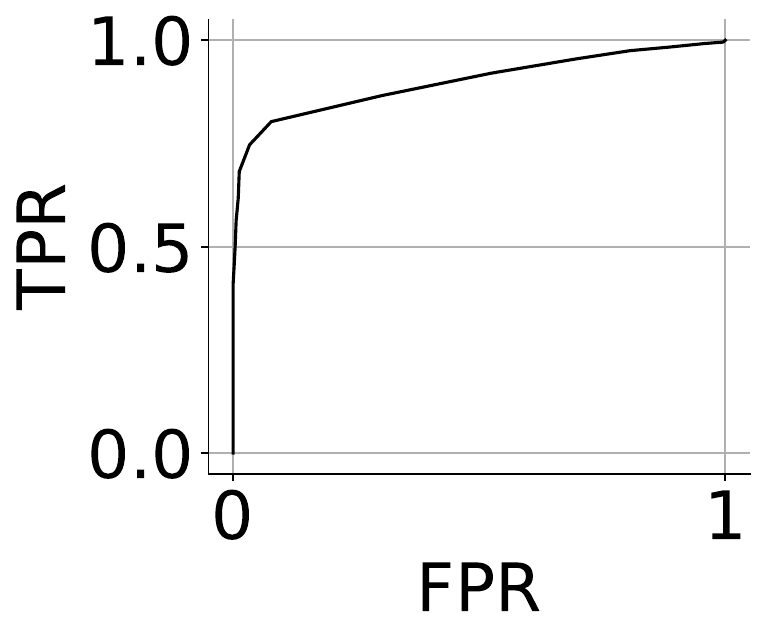}}
  \caption{ROC curves for images taken in static and moving settings.}
\label{fig:roc}
\end{figure}

\section{Conclusion}
In this paper, we present a novel and cost-effective attack to generate robust perturbations in the near-infrared domain, which we dub adversarial infrared perturbations. Our approach ensures real-world robustness by accounting for the spectral shift into the infrared domain and is the first practical attack that works in both targeted and untargeted attack scenarios.
Extensive experiments in the digital and physical domains show that our approach yields consistently high attack success rates in various situations while requiring up to 65\% fewer queries when compared to existing approaches. 
We showed that existing defenses against perturbations cannot successfully defend against our approach. As a remedy, we proposed a novel segmentation-based detection scheme that is specifically designed to thwart our attack with an F1-score of up to $99\%$.

\section*{Acknowledgment}
This work has been co-funded by the Deutsche Forschungsgemeinschaft (DFG, German Research Foundation) under Germany’s Excellence Strategy -EXC 2092 CASA- 390781972, by the German Federal Ministry of Education and Research (BMBF) through the project TRAIN (01IS23027A).

\section*{Ethics Considerations}
Our paper proposes a novel, stealthy, and cost-effective attack to generate both targeted and untargeted robust perturbations. Our measurements have been conducted in a carefully controlled environment to ensure the accuracy and reliability of the data collected. Throughout our measurement process, we took meticulous precautions to guarantee that no harm or disruption occurred to any vehicles or pedestrians on the road. We also ensured at all times full compliance with all applicable safety regulations and standards.

Our primary goal in this paper is to raise awareness about this type of attack and to promote the adoption of suitable defenses for autonomous driving. We explored a range of defensive strategies and demonstrated the feasibility of our proposed countermeasure in mitigating this attack. In this paper, we have only focused on open-source datasets and models and \emph{not} considered any vendor-specific machine learning systems.

We responsibly disclosed our findings—along with our proposed segmentation---based detection schemes---to automotive and camera manufacturers that operate under a system model similar to the one described in \Cref{system_model}, namely Mercedes, Mobileye, Tesla, Sony, and OnSemi.

\bibliographystyle{plain}
\bibliography{infrared_perturbations}

\appendices
\appendices

\section{Infrared Filters and Films}
\label{infrared_films}
Another possible defense would be to embed an infrared spectral filter within the CMOS sensor to protect against attacks that utilize infrared light. While this completely masks the visibility of our infrared perturbations on inputs to the classifier, such an approach is not feasible for several reasons. First, adding spectral filters would significantly impair the ability of cameras to operate in adverse environmental conditions, such as at night or in low-light conditions. In fact, infrared camera vision is instrumental in detecting lane markings and in providing rich contextual information~\cite{10092278}. Second, with over 600,000 electric vehicles sold by Tesla alone in the US~\cite{tesla_sales} in 2023, each containing numerous camera sensors, fitting them all with infrared filters would incur non-negligible extra costs~\cite{wangCanSeeLight2021}. Another approach would be to integrate infrared films within traffic signs. 
In a separate experiment that we performed (cf. \Cref{fig:ir_film}), we observed that this solution considerably hampers the recognition of the traffic sign in the presence of natural ambient light. 

\begin{figure}[tb]
     \centering
     \subfloat[Ambient light.]{\makebox[1.75\width][c]{\centering\includegraphics[height=1.25cm]{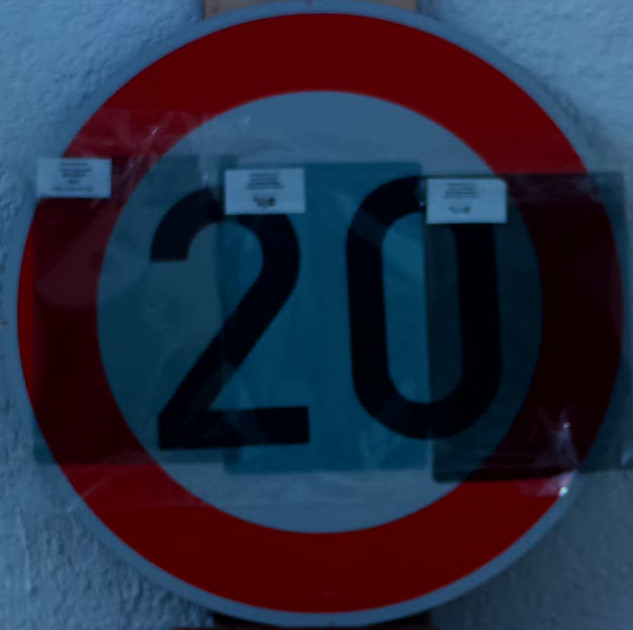}}}
  \subfloat[Infrared light.]{\makebox[1.75\width][c]{\centering\includegraphics[height=1.25cm]{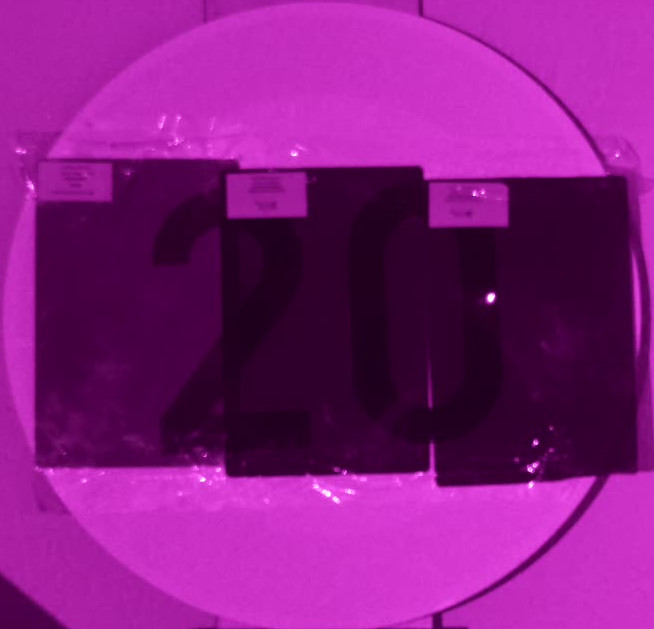}}}
  \subfloat[Ambient/infrared light.]{\makebox[2.5\width][c]{\centering\includegraphics[height=1.25cm]{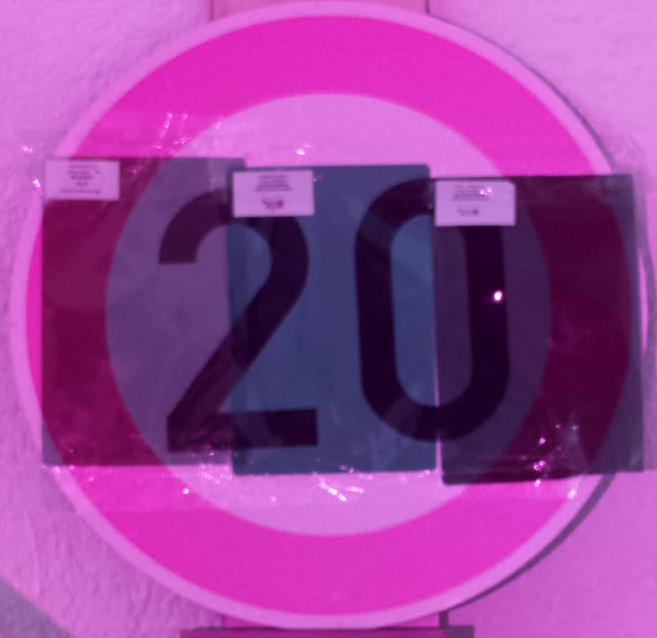}}}
           \caption{Examples of different infrared absorbing films on a speed limit 20 sign with various light sources.}
\label{fig:ir_film}
\end{figure}

\section{IR dataset: GTSRB-IR-100} \label{sec:dataset_ir}

\begin{figure}[tb]
\centering
\includegraphics[width=.5\linewidth]{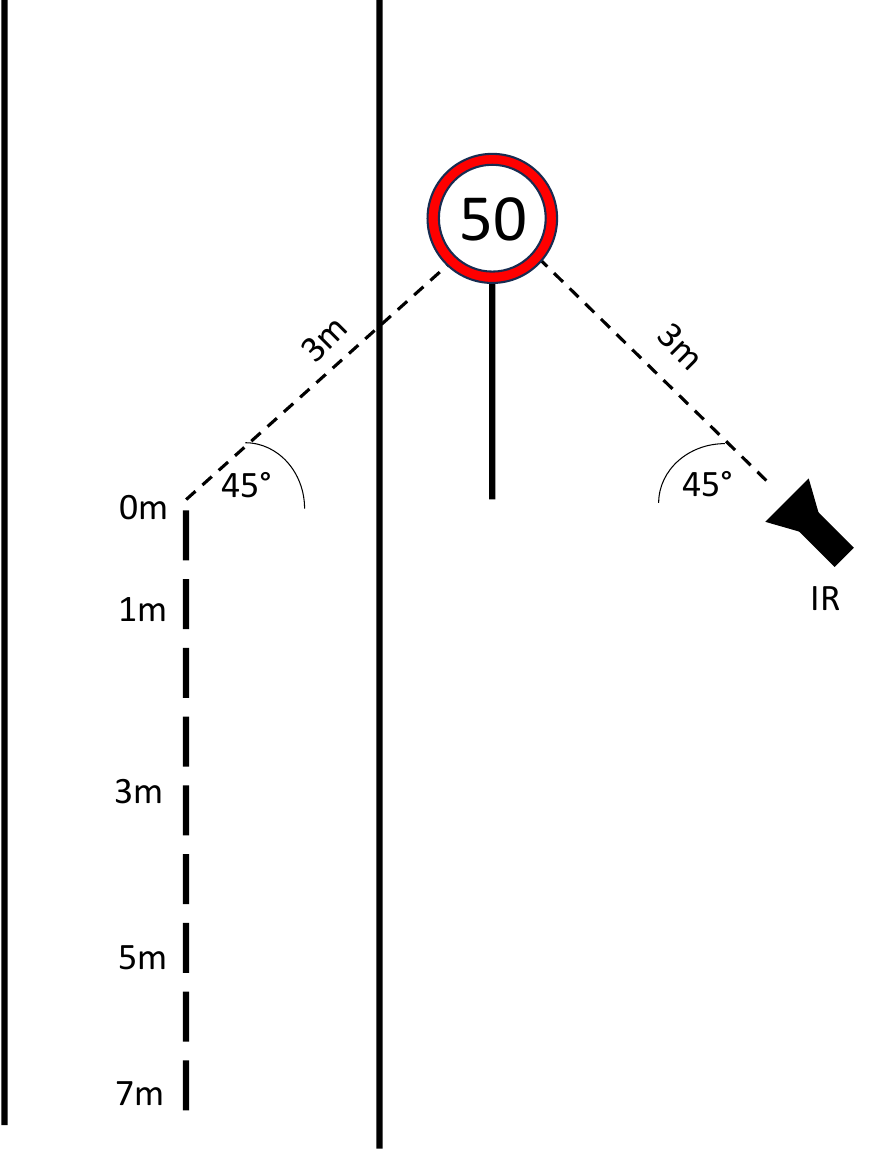}
\caption{Experimental setup used for capturing the \texttt{GTSRB-IR-100} dataset.}
\label{fig:gtsrb-100}
\end{figure}

We publish the dataset \texttt{GTSRB-IR-100}, which comprises 100 images of traffic signs under varying lighting conditions, with half of the images additionally illuminated by an infrared light source. Each image in our dataset is annotated with a lux value measured on the surface of the street sign. To our knowledge, this is the first and only publicly available dataset featuring infrared light sources. 

More concretely, our dataset has been captured according to~\Cref{fig:gtsrb-100} at distances of $\{0, 1, 3, 5, 7\}$ meters. For this, we took an image with ambient light and with an additional infrared light source in the following environments:
\begin{itemize}
    \item Controlled outdoor environment (traffic signs are placed on a stand): Yield, Stop, Speed 20, Speed 50, Speed 100
    \item Realistic-outdoor environment (taking existing traffic signs): Yield, Stop, Turn right, No entry, Speed 30
\end{itemize}

\end{document}